\documentclass[fleqn,usenatbib]{mnras}

\usepackage[T1]{fontenc}

\DeclareRobustCommand{\VAN}[3]{#2}
\let\VANthebibliography\thebibliography
\def\thebibliography{\DeclareRobustCommand{\VAN}[3]{##3}\VANthebibliography}

\usepackage{graphicx}	
\usepackage{amsmath}	
\usepackage{amssymb}	

\usepackage{xcolor}
\usepackage{hyperref}
\newcommand{\orc}{\includegraphics[height=\fontcharht\font`A]{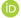}}
\newcommand{\orcid}[1]{\href{https://orcid.org/#1}{\orc}}

\usepackage[normalem]{ulem}

\usepackage{indentfirst}

\usepackage{newtxtext,newtxmath}

\graphicspath{{./}{figures/}}

\def\gsim{\;\rlap{\lower 2.5pt
 \hbox{$\sim$}}\raise 1.5pt\hbox{$>$}\;}
\def\lsim{\;\rlap{\lower 2.5pt
   \hbox{$\sim$}}\raise 1.5pt\hbox{$<$}\;}
\newcommand{\Da}{Damk{\"o}hler }

\title[Molecules and Dust in Galactic Winds]{The Survival and Entrainment of Molecules and Dust in Galactic Winds}

\author[Z. Chen \& S.P. Oh]{
Zirui Chen\orcid{0000-0001-8755-3836}$^{1}$\thanks{E-mail: ziruichen@ucsb.edu} \&
S. Peng Oh\orcid{0000-0002-1013-4657}$^{1}$
\\
$^{1}$Department of Physics, University of California, Santa Barbara, Santa Barbara, CA 93106, USA\\
}

\pubyear{2023}

\begin{document}
\label{firstpage}
\pagerange{\pageref{firstpage}--\pageref{lastpage}}
\maketitle

\begin{abstract}
\noindent Recent years have seen excellent progress in modeling the entrainment of $T \sim 10^4$K atomic gas in galactic winds. However, the entrainment of cool, dusty $T \sim 10-100$ K molecular gas, which is also observed outflowing at high velocity, is much less understood. Such gas, which can be $10^5$ times denser than the hot wind, appears extremely difficult to entrain. We run 3D wind-tunnel simulations with photoionization self-shielding and evolve thermal dust sputtering and growth. Unlike almost all such simulations to date, we do not enforce any artificial temperature floor. We find efficient molecular gas formation and entrainment, as well as dust survival and growth through accretion. Key to this success is the formation of large amounts of $10^4$K atomic gas via mixing, which acts as a protective "bubble wrap" and reduces the cloud overdensity to $\chi \sim 100$. This can be understood from the ratio of the mixing to cooling time. Before entrainment, when shear is large, $t_{\rm mix}/t_{\rm cool} \lsim 1$, and gas cannot cool below the "cooling bottleneck" at $5000$K. Thus, the cloud survival criterion is identical to the well-studied purely atomic case. After entrainment, when shear falls, $t_{\rm mix}/t_{\rm cool} > 1$, and the cloud becomes multi-phase, with comparable molecular and atomic masses. The broad temperature PDF, with abundant gas in the formally unstable $50 \, {\rm K} < T < 5000 \, {\rm K}$ range, agrees with previous ISM simulations with driven turbulence and radiative cooling. Our findings have implications for dusty molecular gas in stellar and AGN outflows, cluster filaments, "jellyfish" galaxies and AGB winds.  
\end{abstract}




\section{Introduction} \label{sec:intro}

Gas flows between a galaxy and its surrounding circumgalactic medium (CGM) regulate the available fuel for star formation. Recent observational advances, exemplified by the COS-Halos survey \citep{Tumlinson:2013}, have allowed us to characterize the diffuse CGM around galaxies \citep{Tumlinson:2017} and in recent years there have been concerted efforts to understand the relevant physical processes \citep{Faucher-Giguere:2023}. One of the most intriguing features of the CGM is its multiphase nature: cold and hot gas with drastically different temperatures coexist on a wide range of scales in the CGM. Yet another prominent CGM feature is the existence of high velocity galactic winds, observed in both the local universe \citep{Lehnert:1996, Martin:1999} and in high-redshift galaxies \citep{Pettini:2001, Shapley:2003}. They play a crucial role in limiting star formation rate and facilitating the gas exchange between galaxies and their CGM. A key question which has received considerable recent attention is how multi-phase gas survives in the presence of a hot wind.

Of particular interest in the multi-phase CGM is the cold, molecular phase, which can directly fuel star formation \citep{Cairns:2019,Colombo:2020} once channeled into the galactic disk. This cold molecular phase often has $T \leq 100$ K and is observed to be outflowing at high velocity, through molecular emission or rotational level transition lines in the Milky Way \citep{Di_Teodoro:2020, Cashman:2021, Noon:2023}, the Large Magellanic Cloud \citep{Tchernyshyov:2022}, low-redshift galaxies \citep{Roussel:2010}, and high-redshift galaxies \citep{Ginolfi:2017}. Additionally, these molecular gas observations are often accompanied by the detection of dust grains. Dust grains are $\sim \mu {\rm m}$ size objects composed of metals such as silicon and carbon; they are crucial in catalyzing molecule formation \citep{Draine:2003}. They are easily sputtered through thermal collisions in high temperature environments \citep{Tsai:1995, Gjergo:2018} but can survive and even grow via accretion \citep{Asano:2013} at cold, molecular temperatures. Dust in the CGM is often detected through reddening \citep{Ménard:2010, Peek:2015}, extinction \citep{Holwerda:2009}, scattering of UV light \citep{Hodges-Kluck:2014}, or thermal emission \citep{Ginolfi:2017}. Molecular gas and dust is also commonly seen in AGN-driven winds \citep{rupke17}. 

The existence of a cold molecular phase at $T \leq 100$ K implies an extreme density contrast of $\gtrsim 10^4$, if it is in pressure equilibrium with the hot, volume filling gas at the virial temperature ($\sim 10^6$ K). It is difficult to understand how such highly overdense gas can become entrained in a tenuous hot wind.

A key problem is that the shear between cold and hot gas should destroy the cloud via Rayleigh-Taylor and Kelvin-Helmholtz instabilities \citep{klein94,scannapieco15,schneider17,zhang17}, on a few cloud-crushing times $t_{\rm cc}$. Yet, cold gas is seen in abundance outflowing in galactic winds, and also infalling as HVCs. Recent progress in understanding atomic ($T \sim 10^4$K) gas entrainment highlights the role of radiative cooling, or condensation \citep{armillotta17,GronkeOh:2018,gronke20-cloud,li20,schneider20,kanjilal21,Abruzzo:2022,TanFielding:2023}. Hydrodynamic instabilities cause the cold gas and hot gas to mix. If the mixed gas cools faster than it is produced, the cloud will survive and grow. For this to happen, the cooling time of mixed gas $t_{\rm cool,mix}$ must be shorter than the cloud crushing time $t_{\rm cc} \sim \sqrt{\chi} r_{\rm cloud}/v_{\rm wind}$, where $r_{\rm cloud}$ is the initial cloud size, $v_{\rm wind}$ wind is the wind velocity, and $\chi = \rho_{\rm c}/\rho_{\rm h}$ is the overdensity of cold cloud gas. This implies that the cloud must be larger than a critical scale $r > r_{\rm crit} \sim v_{\rm wind} t_{\rm cool,mix}/\sqrt{\chi}$ \citep{GronkeOh:2018}. The cooled gas retains its initial momentum, and in time cold gas both grows in mass and comoves with the hot gas. This form of mixing-induced thermal instability enforces kinematic coupling between phases, and gas is converted from hot to cold. Turbulent mixing layers (TMLs) are generally the primary interface where different phases exchange mass, momentum and energy, and in recent years there has been substantial progress on our understanding of their properties \citep{begelman90,kwak10,ji19,Fielding:2020,tan21,chen23}.

However, almost all cloud-crushing simulations to date focussed on optically thin, $T \sim 10^4$ K atomic clouds. These simulations have a temperature floor set at $T \sim 10^4$ K to mimic the heating effect of the photo-ionizing UV background. This implies a typical density contrast of $\chi \sim 100-1000$ for clouds in a $T \sim 10^6-10^7$K wind. The survival and entrainment of $T \sim 10-100$K dusty, molecular clouds with density contrasts of $\chi \sim 10^4-10^5$ is much less well understood, particularly given the different cooling curve at low temperatures and the sensitivity of dust to destruction by sputtering when it mixes with hot gas. Computationally, the molecular cloud problem is thought to be much more challenging, since typically clouds form a cometary tail of length $\sim \chi r_{\rm cloud}$, where most of the mass and momentum transfer takes place. Thus, studying molecular clouds in principle requires boxes $\sim 10^4-10^5$ times longer then the initial cloud size. Since the initial cloud has to be resolved by at least $\sim 8$ cells \citep{GronkeOh:2018}, this is extremely challenging in 3D simulations.

To fill in this gap, \cite{farber22} performed a pioneering study of the survival of dusty clouds in wind tunnel simulations. To make the problem computationally tractable, they enforced an artificial temperature floor of $T \sim10^3$ K. They find three possible pathways for $T \sim 10^3$K clouds in a hot $T \sim 10^6$K wind: in order of increasing initial cloud size, the clouds can be destroyed, transformed to $T \sim 10^4$K gas, or survive with substantial $T \sim 10^3$K gas. By modeling dust via Lagrangian tracer particles, they also find that dust can in principle survive, depending on the amount of time it spends in the hot wind. \cite{Girichidis:2021} run similar run wind tunnel simulations. While they do not demonstrate long-term cloud entrainment and survival, as the system is evolved for only a few cloud-crushing times, they include many more physical processes, including non-equilibrium chemistry, self-gravity, magnetic fields, as well as photoionization and self-shielding, and do not enforce a temperature floor. They focus on the possibility of {\it in situ} formation of molecular gas in initially warm ($T\sim  10^3-10^4$K) clouds, and find an abundance of warm gas in their simulations, themes which we will revisit in this work. 

This paper is motivated by observations that show that cold molecular gas in outflows is never found in isolation, but is always accompanied by a comparable mass of warm, ionized atomic gas \citep{Rupke:2018,Veilleux2005}. An abundance of warm gas is also seen in the cloud-crushing simulations described above \citep{Girichidis:2021,farber22}, though its origin and implications merit further exploration. A minimal amount of warm atomic gas arises from photoionization, before self-shielding kicks in. In addition, turbulent mixing typically produces gas with temperatures of order the geometric mean of the hot and cold phases $T_{\rm mix} \sim (T_c T_h)^{1/2}$ \citep{begelman90}. For $T_{\rm c} \sim 10^4$K clouds in a $T_{\rm h} \sim 10^6$K wind, $T_{\rm mix} \sim 10^5$K, which is highly thermally unstable. However, for $T_{\rm c} \sim 10-100$K molecular gas, $T_{\rm mix} \sim 10^{3.5}-10^{4}$K, which is thermally stable and will survive for a long time, potentially as part of a three-phase medium. We will explore these ideas. The existence of warm atomic gas alongside molecular gas has important consequences for the survival and entrainment of molecules and dust. First, atomic gas invariably arises at the surface of a cloud (e.g., from photoionization), and serves as protective "bubble wrap" which prevents molecules and dust from directly interacting with the hot phase and getting destroyed. Second, atomic gas greatly decreases the overdensity of a multi-phase gas cloud. For instance, if there are comparable amounts of $T \sim 10^4$K and $T\sim 10$K gas in a cloud, if all phases are in pressure balance the overdensity will be $\chi \sim 200$ (i.e., the cloud has roughly the same volume as a purely atomic cloud, but has double the mass), not $\chi \sim 10^4 - 10^5$. This makes entrainment and survival essentially equivalent to that of a purely atomic cloud, which is already well understood and computationally tractable. We therefore study the long-term survival and entrainment of multi-phase gas where we consider the full dynamic range of the cooling curve without any temperature floor (i.e., gas can cool down to $T \sim 10$K), a regime which has not been probed in numerical simulations. 

The outline of this paper is as follows. In \autoref{sec:methods}, we introduce our simulation setup and our implementation of processes such as photoionization, self-shielding, and dust evolution. In \autoref{sec:results}, we present results of a suite of simulations of three different cloud sizes, focusing on the survival and entrainment of molecular-temperature gas and dust. In \autoref{sec:understading molecular cloud entrainment}, we delve into the physical reasons behind the plethora of warm gas we find, by running simulations where we change photoionization and cooling curve properties. In \autoref{sec:discussion}, we discuss the implications of our work. Finally, in \autoref{sec:summary}, we summarize.

\section{Methods} \label{sec:methods}
We run our simulations with Athena++ \citep{Stone:2020}, which solves the 3D hydrodynamical equations on a uniform Cartesian grid. Our initial conditions consists of a stationary spherical cloud at $T \sim 10^4 {\rm K}$ placed in a rectangular box filled with a hot wind at $10^6 {\rm K}$ with Mach number $\mathcal{M}_{\rm wind} = 1.5$, which mimics the transonic outflows seen in galactic winds. Our fiducial runs also have hot and cold gas number densities of $n_{\rm h} \sim 10^{-3} \, {\rm cm^{-3}}$, $n_{\rm c} \sim 0.1 \, {\rm cm^{-3}}$ respectively. To avoid numerical artifacts, we randomly seed percent-level density fluctuations in the initial cold cloud. This cloud-in-wind setup is initialized to be in pressure equilibrium and placed in a rectangular box with dimensions $(\sim 10 r_{\rm initial})^2 \times (\sim 120 r_{\rm initial})$, where $r_{\rm initial}$ is the radius of the initial cold cloud, and the box size in the direction parallel to the wind is extended such that we capture the tail formation of the cloud. The boundary conditions of the simulation box are all outflowing, except that we impose a constant inflow of hot gas in the wind direction. Similar to \cite{McCourt:2015}, \cite{GronkeOh:2018} and \cite{tan23-gravity}, we adopt a cloud-tracking scheme that continuously shifts to the center-of-momentum frame of the cloud. This scheme allows us to track the cloud as it gets accelerated and entrained in the hot wind. By using a cloud tracking scheme in our simulations, we minimize the relative velocity between the cloud and the computation grid, which reduces truncation errors and allows for faster numerical convergence \citep{McCourt:2015}. This also helps us avoid issues related to preserving Galilean invariance when using a grid-based code \citep{Tasker:2008}. Furthermore, \cite{Li_Zhihui:2019} conducted cloud-crushing simulations using the Lagrangian, mesh-free code GIZMO \citep{Hopkins:2015} and obtained results that are consistent with simulations performed with grid codes. Our simulation setup is therefore robust.

Besides this standard wind tunnel setup, we also implement radiative cooling, self-shielding, and dust sputtering/accretion, which are described in the following sections. We also provide a summary of all the simulations we run in \autoref{sec:summary of simulations we run} and \autoref{tab:summary of simulations}.

\subsection{Radiative Cooling} \label{sec:radiative cooling}

The net cooling function $\mathcal{L}$ is given by
\begin{align}
    \rho \mathcal{L} = n^2 \Lambda - n \Gamma
\end{align}
where $\Lambda(T)$ is the cooling function, and $\Gamma$ is the heating rate. For the cooling function $\Lambda(T)$, we implement the module developed by \cite{TanFielding:2023}, which combines the collisional ionization equilibrium (CIE) cooling curve in \cite{Gnat:2007} for $T \geq 10^4 {\rm K}$ with the cooling function in \cite{Koyama:2002} for $T \leq 10^4 {\rm K}$ and assumes solar metallicity (${\rm X}=0.7$, ${\rm Z}=0.02$). The cooling curve is obtained by performing a piece-wise power law fit over $\sim 50$ logarithmically spaced temperature bins from $10 {\rm K}$ to $10^{10} {\rm K}$. This cooling curve is implemented using the fast and robust exact cooling algorithm described in \cite{Townsend:2009}. In our simulations, we shut off cooling above 0.6 times the wind temperature to prevent wind cooling. For the heating rate $\Gamma$, we impose a temperature floor at $10^4 {\rm K}$ for photoionized gas (i.e., gas which is optically thin and not yet self-shielding) to mimic photo-heating. 

\begin{figure}
\centering
\includegraphics[width=\columnwidth]{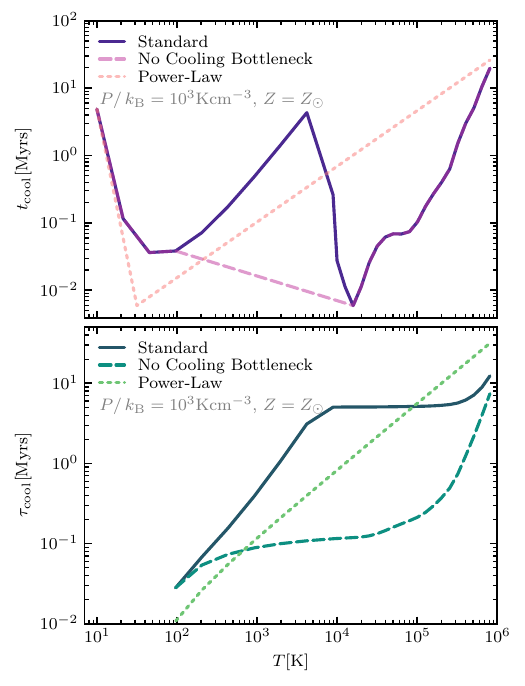}
\caption{In the top panel, we show the standard isobaric cooling time $t_{\rm cool}$ (purple, obtained using \autoref{eq:t_cool}) vs. temperature with $\left. P \right/ k_{\rm B}=10^3 {\rm K} {\rm cm}^{-3}$ and solar metallicity. We consider the full dynamic range of temperatures from 10K to $10^6$K in our simulations. Note the sharp rise in the standard $t_{\rm cool}$ profile just below $T \sim 10^4$K due to hydrogen recombination. We will refer to this feature as the "cooling bottleneck" and will show that it plays an important role in regulating the abundance of $T \sim 10^4$K gas. We also include two variations of the standard $t_{\rm cool}$ profile. In the "No Cooling Bottleneck" profile (magenta), we take out the cooling bottleneck just below $T \sim 10^4$K by connecting the cooling times between $\sim$ 200K and $\sim 2 \times 10^4$K using a power-law. We will use this modified profile to explore how the cooling bottleneck affects cloud composition and survival outcomes in \autoref{sec:What Role Does Photo-ionization and the Cooling Bottleneck Play?}. In the "Power-Law" case (pink), we adopt a two-piece power-law cooling time profile that is intended to mimic the $10^4$K to $10^6$K cooling time profile over a broader temperature range between 10K and $10^6$K. We describe the exact procedure of obtaining this power-law profile in \autoref{sec:Survival Criterion for Clouds with Molecules} and use it to gain insights about the survival criterion for molecular-temperature clouds there. In the bottom panel, we show profiles of $\tau_{\rm cool}$, the time needed to cool down to the minimum of the isobaric cooling time $t_{\rm cool}$ in the molecular temperature range at $T = 45$K, as a function of temperature for each of the three $t_{\rm cool}$ profiles we considered in the top panel. $\tau_{\rm cool}$ can be obtained from $t_{\rm cool}$ through \autoref{eq:tau_cool}. In \autoref{sec:The Damkohler Number and Atomic Gas Production Through Mixing}, we will use $\tau_{\rm cool}$ when we compare the characteristic mixing and cooling timescales of the cloud during entrainment. Note that $\tau_{\rm cool}$ is set by the cooling time at the bottleneck for $5000 K < T < 5 \times 10^5$K, and corresponds to the standard cooling time for $T < 5000$K.}
\label{fig:cooling_curve}
\end{figure}

In the top panel of \autoref{fig:cooling_curve}, we plot the standard isobaric cooling time $t_{\rm cool}$ (purple) as a function of temperature from 10K to $10^6$K, which is the temperature range we are primarily interested in. The isobaric cooling time is: 

\begin{align}
    t_{\rm cool} \sim \frac{P}{n^2 \Lambda(T)} \sim \frac{P}{{(\frac{P}{k_B T})}^2 \Lambda(T)} \sim \frac{k_B^2 T^2}{P \Lambda(T)}. \label{eq:t_cool}
\end{align}

Note the sharp rise in cooling time below $T \sim 10^4$K due to hydrogen recombination. We will find that this "cooling bottleneck" plays an important role in regulating the abundance of $T \sim 10^4$K gas. The minima in $t_{\rm cool}$ at $T \sim 50, 10^4$K are associated (in the presence of heating) with the stable phases corresponding to cold molecular-temperature and warm atomic gas respectively. 

In the top panel of \autoref{fig:cooling_curve}, we also include two modified $t_{\rm cool}$ profiles (we refer to them as the "No Cooling Bottleneck" profile and the "Power-Law" profile) which will be used in controlled numerical experiments in \autoref{sec:understading molecular cloud entrainment} to help us better understand the entrainment and survival criterion of molecular-temperature clouds. We briefly describe these modified profiles in the caption of \autoref{fig:cooling_curve} but direct the readers to \autoref{sec:understading molecular cloud entrainment} for a much more detailed discussion of the motivation and implementation of these modified profiles.

In the bottom panel of \autoref{fig:cooling_curve}, we show profiles of $\tau_{\rm cool}$, the time needed to cool down to the minimum of the isobaric cooling time $t_{\rm cool}$ in the molecular temperature range at $T = 45$K\footnote{We choose the temperature $T \sim 45$K where $t_{\rm cool}$ has a minimum instead of the temperature floor at $T \sim 10\,$K, because $t_{\rm cool}$ rises steeply from $T \sim 45\,$K to 10 K. Choosing a lower limit of $T \sim 10$ K would overestimate cooling times, just as choosing a lower limit of $T \sim 5000$K instead of the minimum cooling time at $T \sim 1.5 \times 10^4 \,$K would overestimate the cooling time down to atomic temperatures. For cosmetic purposes like making slice plots for $\tau_{\rm cool}$ in our simulations, we simply set $\tau_{\rm cool}\left( T \leq 45{\rm K}\right)$ = $t_{\rm cool}\left( 45{\rm K}\right)$. Our conclusions about cooling down to low molecular temperatures are unaffected by the latter choice.}, as a function of temperature for each of the three $t_{\rm cool}$ profiles we considered in the top panel. $\tau_{\rm cool}$ can be obtained from $t_{\rm cool}$ using

\begin{align}
    \tau_{\rm cool}=  \int_{45 {\rm K}}^{T_{\rm current}} \frac{t_{\rm cool}(T)}{T} \,dT. \label{eq:tau_cool}
\end{align}

In \autoref{sec:The Damkohler Number and Atomic Gas Production Through Mixing}, we will use $\tau_{\rm cool}$ when we compare the characteristic mixing and cooling timescales of the cloud during entrainment.

\subsection{Photoionization and Self-Shielding} \label{sec:self-shielding}

We treat photoionization and self-shielding in simple and heuristic manner. In particular, we do {\it not} perform radiative transfer calculations. We simply assume that the cloud is surrounded by a photoionized "skin", where we set the temperature floor to be at $T \sim 10^4$K. Once the cloud starts to self-shield and become optically thick, then we remove this temperature floor. The critical HI column density for self-shielding is $N_{\rm HI} \sim \sigma_{\rm HI}^{-1} \sim 10^{17} \, {\rm cm^{-2}}$, where $\sigma_{\rm HI}$ is the hydrogen photoionization cross-section. Assuming the neutral fraction of the optically thin region to be $x_{\rm HI} \sim 1\%$ (we discuss this choice further below), the total hydrogen column density of the cloud "skin" is: 
\begin{equation}
      N_{\rm skin} \sim N_{\rm HI} x_{\rm HI}^{-1} \sim 10^{19} {\rm cm}^{-2} \left(\frac{x_{\rm HI}}{0.01}\right)^{-1} , \label{eq:self-shielding column}
\end{equation}
or
\begin{align}
     d_{\rm skin} \sim \frac{N_{\rm skin}}{n_{\rm atomic}} \sim 30 \, {\rm pc} \, n_{\rm atomic,-1}^{-1} \left(\frac{x_{\rm HI}}{0.01}\right)^{-1}  \label{eq:cloud skin depth}
\end{align}
where $n_{\rm atomic,-1} \equiv \left. n_{\rm atomic} \right/ 10^{-1} \: {\rm cm}^{-3}$.

For each cloud cell, which we define to be cells with $T < 10^{4.5} {\rm K}$, we look through all the cells with distance $\leq d_{\rm skin}$ from it. If any of these neighboring cells have $T > 10^{4.5} {\rm K}$, then this cloud cell is not self-shielded. Since it is photoionized by the ambient UV background, we set the temperature floor for this cloud cell at $T = 10^4 {\rm K}$. Otherwise, if all the neighboring cells within $d_{\rm skin}$ satisfy  $T < 10^{4.5} {\rm K}$, the cloud cell is deemed to be self-shielded and allowed to cool down to molecular temperatures. For such cells, we set the temperature floor at $T=10$ K, which is where our cooling function described in \autoref{sec:radiative cooling} terminates.

For our fiducial value of $\langle x_{\rm HI} \rangle \sim 1\%$, we follow \citet{Werk:2014}, which used CLOUDY \citep{Ferland:1998, Ferland:2013} to model the CGM of 33 COS-Halos galaxies. With the \cite{HM:2001} background radiation field, they vary the dimensionless ionization parameter to fit for the observed column densities, and find $\langle x_{\rm HI} \rangle \sim 1\%$. They also checked the effect of ionizing radiation from the central galaxy and found it to be relatively unimportant. Note that $x_{\rm HI}$ depends on the ionization parameter, i.e. both the ambient radiation field and the cloud density (which depends on the pressure of the surrounding medium), so it should be expected to evolve with radius, virial temperature and redshift. We therefore treat $x_{\rm HI}$ as a free parameter, though in practice our results are not sensitive to this assumption.

Another potentially relevant quantity here is the column density required to self-shield molecular hydrogen \citep{li20}. This has been well-studied in the context of the ISM and corresponds to a surface density of $ 10 \, M_{\odot} \, {\rm pc}^{-2}$ or $N_{\rm H} \sim 10^{21} {\rm cm}^{-2}$ at solar metallicity \citep{Robertson:2008,Krumholz:2011}. Here, we do not model the formation of molecular hydrogen. We are primarily focused on temperature evolution. At solar metallicity, fine-structure metal-line cooling dominates and brings gas to cold temperatures ($T < 100$K) without requiring molecular line cooling \citep{Krumholz:2012,Glover:2013}. Of course, we do expect dust catalyzed molecule formation \citep{Wakelam:2017,Grieco:2023}, consistent with observational signatures such as ${\rm CO}$ emission lines and ${\rm H}_2$ rotational level transition lines, but explicitly modeling this is beyond the scope of this paper. In the remainder of this paper, we will refer to the phase below $\sim 3000$K as "molecular-temperature gas" or the "molecular-temperature phase", with the understanding that although we do not explicitly model the formation of molecules, we do expect dust catalyzed molecule formation to proceed at these temperatures.

\subsection{Dust Sputtering and Accretion} \label{sec:dust sputtering and accretion}
As discussed in \autoref{sec:intro}, a crucial constraint on cloud models is the presence of dust at large circumgalactic radii. Since dust cannot survive in hot gas, it must be preserved and/or regenerated during cold gas transport from the galactic ISM to the CGM. As dust is exposed to the hot wind in turbulent mixing layers, it can be sputtered away by thermal collisions. The temperature and density dependent dust sputtering time is \citep{Tsai:1995, Hirashita:2018, Gjergo:2018}

\begin{align}
     t_{\rm sputter} = 70 {\rm Myrs} \left( \frac{a} {0.1 \, \mu m} \right) \left( \frac{n_{\rm H}}{10^{-3} {\rm cm^{-3}}} \right)^{-1}    \left[1+ \left(\frac{T} {2 \times 10^6 {\rm K}}\right)^{-2.5}\right].
     \label{eq:dust sputtering time}
\end{align}

where $a$ is the grain size. For simplicity, we shall assume a single grain-size population, given by the typical average grain size $\langle a \rangle = 0.1 \mu$m \citep{mathis77}. Besides sputtering, dust can also experience mass growth through accretion \citep{Liffman:1989, Draine:2009, Jones:2011} in a cold, high density environment. The dust accretion timescale is \citep{Asano:2013}

\begin{align}
     t_{\rm acc} = 200 \, {\rm Myr} \, \left( \frac{a} {0.1 \, \mu m} \right) {\left(\frac{n_{\rm H}}{20 \, {\rm cm}^{-3}}\right)}^{-1}   {\left(\frac{T}{50 {\rm K}}\right)}^{-\frac{1}{2}} \left( \frac{Z}{Z_{\odot}} \right)
     \label{eq:dust accretion time}
\end{align}

assuming spherical dust grains. Note that this assumes that all molecules or atoms which collide with a dust grain stick to it (i.e., a sticking coefficient of 1), which can be inaccurate at higher temperatures $T \gsim 10^4$ K, or when the dust grains are exposed to strong UV radiation fields (e.g. in AGN winds). However, we expect the dust accretion in our simulations to be dominated by the low temperature $T < 100 {\rm K}$ phase in the self-shielded cloud interior. Also, note that dust growth cannot continue indefinitely but must saturate at a value commensurate with the gas metallicity -- for instance, a highly supersolar dust mass fraction cannot arise in solar metallicity gas. Similar to \citet{hirashita17},  we enforce this by using a dust growth time: 

\begin{align}
    t_{\rm grow} = \frac{t_{\rm acc}}{1-{\mathcal D}/{\mathcal D}_{\rm sat}} \label{eq:dust grow time} 
\end{align} 

where $\mathcal{D}=\rho_{\rm dust}/\rho_{\rm gas}$ is the dust mass fraction and $\mathcal{D}_{\rm sat}$ is the saturated dust fraction. Since we assume solar abundances, for us $\mathcal{D}_{\rm sat} \sim Z_{\odot}$. 
Note that we have scaled densities and temperatures to values appropriate for our fiducial runs, and that the dust evolution timescales are considerably longer than typical cooling times.

To study dust evolution in our simulations, we model dust by a passive scalar $\rho_{\rm dust} = \mathcal{D} \rho_{\rm gas}$ and simply evolve it in each simulation cell according to the equation
\begin{equation}
\frac{d \rho_{\rm dust}}{dt} = \frac{\partial \rho_{\rm dust}}{\partial t}  + \nabla \cdot (\rho_{\rm dust} \mathbf{v}) =  \frac{\rho_{\rm dust}}{t_{\rm grow}} - \frac{\rho_{\rm dust}}{t_{\rm sputter}}
\end{equation}
where we use $t_{\rm sputter}(n,T)$, $t_{\rm grow}(n,T)$ from equations \ref{eq:dust sputtering time} and \ref{eq:dust grow time} respectively. We therefore assume that dust is strongly coupled and simply advects with the gas; it does not diffuse relative to the gas. Dust updates are performed during simulation run time, and not in post processing. This is a fairly crude treatment, but it suffices to make physical estimates for dust survival.

\subsection{Resolution} \label{sec:resolution}
Cloud-crushing simulations typically require at least $r_{\rm initial}/d_{\rm cell} \sim 8$ (where $r_{\rm initial},d_{\rm cell}$ are the initial cloud size and the cell size respectively) for converged mass growth and cloud entrainment \citep{GronkeOh:2018}. However, our simulations with photoionization impose an additional constraint: the optically thin "skin" of the clouds must be numerically resolved. In \autoref{sec:results}, we will analyze three representative clouds with initial sizes $r_{\rm initial} = 3, 10, $ and 30 $r_{\rm crit}$ (or $100{\rm pc}$, $300 {\rm pc}$, and $1 {\rm kpc}$ in physical units). For these three clouds, we choose our fiducial resolution to be $\left. r_{\rm initial}\right/d_{\rm cell} = $ 15, 30, and 60, respectively, when the cloud is at $T \sim 10^4$K. This corresponds to resolving the optically thin cloud skin by $5$, $3$, and $2$ cells. While this means that the skin is only marginally resolved in fiducial runs, we demonstrate reasonable convergence in cloud mass evolution in \autoref{sec:convergence test}. 

\subsection{Summary of Simulations We Run} \label{sec:summary of simulations we run}

\begin{table*}
\caption{Summary of the Simulations We Run}
\centering
\begin{tabular}{c c c c c c c c} 
\hline\hline 

\vtop{\hbox{\strut $\left.r \right/r_{\rm crit}$}} &
\vtop{\hbox{\strut $T_{\rm cloud}$ [K]}} & 
\vtop{\hbox{\strut $T_{\rm wind}$ [K]}} & \vline &
\vtop{\hbox{\strut Photoionization}} &
\vtop{\hbox{\strut Dust Sputtering \& Accretion}} &
\vtop{\hbox{\strut Cooling}} &
\vtop{\hbox{\strut $\left.r_{\rm cloud} \right/d_{\rm cell}$}} \\ 
\hline 
3 & $10^4$ & $10^6$ & \vline & Yes & Yes & Standard & 15 \\
10 & $10^4$ & $10^6$ & \vline & Yes & Yes & Standard & 30\\
30 & $10^4$ & $10^6$ & \vline & Yes & Yes & Standard & 60 \\
\hline
10 & $10^4$ & $10^6$ & \vline & No & Yes & Standard & 30\\
10 & $10^4$ & $10^6$ & \vline & Yes & Yes & No Cooling Bottleneck & 30\\
10 & $10^4$ & $10^6$ & \vline & No & Yes & No Cooling Bottleneck & 30\\
\hline
10 & $10^4$ & $10^6$ & \vline & Yes & Yes & Power-law & 30\\
\hline
static cloud & $10^2$ & $10^6$ & \vline & Yes & Yes & Standard & 7.5\\
\hline
3 & $10^4$ & $10^7$ & \vline & No & Yes & Standard & 15 \\

\hline\hline
\end{tabular}
\label{tab:summary of simulations}
\end{table*}

Our fiducial suite of simulations implements all the physics described above while varying the initial cloud radius. We will explain and motivate our choices of the initial cloud radii in \autoref{sec:results}. To understand how the shape of the cooling curve and photoionization affects cloud evolution, we run a number of additional simulations where we modify these properties. We have briefly mentioned these modifications in \autoref{sec:radiative cooling} but will provide a much more detailed analysis and present the insights we gained in \autoref{sec:understading molecular cloud entrainment}. We also simulate a static cloud, which we discuss in \autoref{sec:The Damkohler Number and Atomic Gas Production Through Mixing}, to better understand the composition of the entrained cloud in our fiducial setup. Finally, we run an additional simulation with $T_{\rm wind}=10^7$ K to explore the effects of a hotter wind temperature. We summarize all the simulations we run in \autoref{tab:summary of simulations}.

\section{Results}\label{sec:results}

In this section, we present results from our simulations described in \autoref{sec:methods}. We normalize timescales to the cloud-crushing time, which is a typical timescale for cloud destruction in adiabatic simulations \citep{klein94}: 
\begin{equation}
t_{\rm cc} = \chi^{1/2} \frac{r_{\rm initial}}{v_{\rm wind}} \sim 13 \, {\rm Myr} \left( \frac{\chi}{100} \right)^{1/2} \left( \frac{r_{\rm initial}}{200 \, {\rm pc}} \right) \left( \frac{v_{\rm wind}}{225 \, {\rm km \, s^{-1}}}  \right)^{-1} 
\end{equation}

We also need to define a critical radius for cloud survival. All our simulations are scaled to this radius. In general, this requires that a mixing time $t_{\rm cc} \propto \chi^{1/2} r/v$ (where $\chi = \rho_{\rm c}/\rho_{\rm h}$ is the cloud overdensity) be longer than a cooling time $t_{\rm cool}$, so that there is a critical radius for cloud survival $r > r_{\rm crit} \sim v t_{\rm cool}$ \citep{GronkeOh:2018}. However, as a long list of papers on this topic attest \citep{GronkeOh:2018,sparre19,li20,kanjilal21,Abruzzo:2022,farber22,Abruzzo:2023}, {\it which} particular velocity and {\it which} particular cooling time is appropriate has proven controversial, and in particular sensitive to the adopted temperature floor for the cloud, as well as the overdensity and Mach number. We discuss this further in \autoref{sec:Survival Criterion for Clouds with Molecules}.

Regardless of the physical reasoning behind it, for any cooling curve and temperature floor, there {\it is} a critical cloud radius for survival, which can be obtained by running a suite of numerical simulations that probe the border between cloud growth and destruction. For a $T \sim 10^4$K temperature floor\footnote{Originally, a $T\sim 10^4$K temperature floor was thought to arise due to photoionization. In this paper, we show that an effective $T\sim 10^4$K temperature floor also arises during the entrainment process due to mixing.}, and other parameters (cloud overdensity, wind speed, etc) appropriate to our simulations, the critical radius is \citep{Abruzzo:2023} : 
\begin{align}
    r_{\rm crit} 
    \sim 30\, {\rm pc} \, \frac{T_{\rm cl,4}^{\left. 3\right/2}}{P_{\rm 3}\Lambda_{\rm mix,-21.4}} \left(\frac{\mathcal{M}_{\rm wind}}{1.5} \right) \left(\frac{\chi}{100}\right)^{\left. 3\right/2} \left( \frac{\alpha}{7}\right)^{-1}, \label{eq:r_crit abruzzo}
\end{align}
where $T_{\rm cl,4} \equiv \left. T_{\rm cl} \right/ 10^4 \: {\rm K}$, $P_{\rm 3} \equiv \left. nT \right/ \left(10^3 {\rm cm}^{-3} \: {\rm K}\right)$, $\Lambda_{\rm mix,-21.4} \equiv  \left. \Lambda\left(T_{\rm mix}\right) \right/ \left(10^{-21.4} \: {\rm erg} \: {\rm cm}^3 \: {\rm s}^{-1}\right)$, $\mathcal{M}_{\rm wind}$ is the Mach number of the hot wind, and $\alpha$ is a constant that is calibrated by numerical simulations. Surprisingly, equation \ref{eq:r_crit abruzzo} turns out to be appropriate even for clouds which can cool down to $T\sim 10$K. As we shall see, this is because they remain mostly atomic during the entrainment process. 

A crucial dimensionless ratio that determines the percentage of the optically thin "skin" in the cloud is $N_{\rm crit}/N_{\rm skin}$. Equation \ref{eq:r_crit abruzzo} corresponds to a column density: 
\begin{align}
    N_{\rm crit} \sim 10^{19} {\rm cm}^{-2} n_{-1} \frac{T_{\rm cl,4}^{\left. 3\right/2}}{P_{\rm 3} \Lambda_{\rm mix,-21.4}} \left(\frac{\mathcal{M}_{\rm wind}}{1.5} \right) \left(\frac{\chi}{100}\right)^{\left. 3\right/2} \left( \frac{\alpha}{7}\right)^{-1}. \label{eq:N_crit abruzzo}
\end{align}
Combining \autoref{eq:N_crit abruzzo} and \autoref{eq:self-shielding column}, we find that for our parameter choices, 
\begin{align}
    N_{\rm crit}/N_{\rm skin} \sim 1. \label{eq:r_crit over d_skin}
\end{align}
Thus, clouds of size $r_{\rm crit}$ have photoionization optical depths of order unity, while larger clouds are self-shielding. We therefore focus on clouds which are several times $r_{\rm crit}$, and thus able both to survive and self-shield in order to cool to low temperatures\footnote{Also note that as the cloud is sheared by the wind, it becomes elongated, reducing the tranverse column density from its initial value. Although a $N_{\rm initial} = N_{\rm crit}$ cloud is able to survive, get entrained, and eventually become optically thick to produce molecular-temperature gas, we expect this process to take a long time.}.

In the following sections, we analyze the survival and evolution of molecular-temperature gas and dust in our clouds. We focus on a $r_{\rm initial} = 10 r_{\rm crit}$ cloud in \autoref{sec:intermediate cloud} but also discuss smaller and larger clouds with $r_{\rm initial} = 3$ and $10 r_{\rm crit}$ respectively.

\subsection{A Cloud with $r_{\rm initial} = 10 r_{\rm crit}$} \label{sec:intermediate cloud}

\begin{figure*}
\centering
\includegraphics[width=\textwidth]{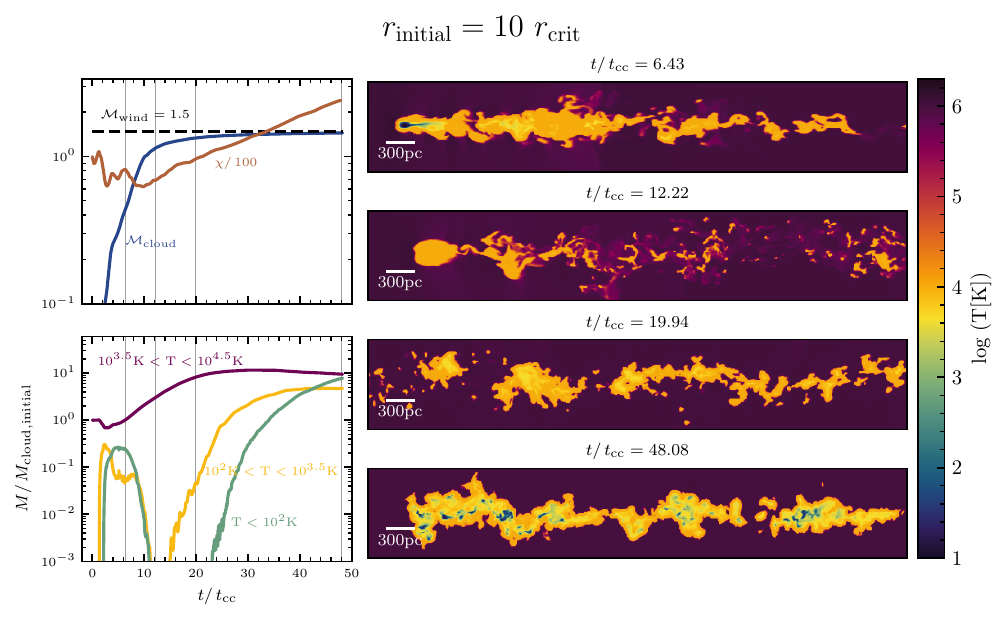}
\caption{Evolution of a cloud with initial size $r_{\rm initial} = 10 r_{\rm crit}$. The initial conditions and simulation details are described in \autoref{sec:methods}. The top left panel plots the time evolution of three important dimensionless numbers: the Mach number of the wind (dashed black, $\left. v_{\rm wind} \right/ c_{\rm s,hot}$), the Mach number of the cloud (blue, $\left. v_{\rm cloud} \right/ c_{\rm s,hot}$, where $v_{\rm cloud}$ is the center of momentum velocity of the cloud), and the average density contrast (brown, $\left. \overline{\rho}_{\rm cloud} \right/ \rho_{\rm wind}$) The Mach number of the cloud asymptotically approaches the Mach number of the wind as the cloud gets entrained.The average density contrast never exceeds 100 in the first 20 $t_{\rm cc}$, which means the cloud is mostly atomic during entrainment. This is why the survival criterion for optically thin atomic clouds work well in our setup. The abundance of atomic gas facilitates cloud survival and allows it to be entrained at $\left.t\right/ t_{\rm cc} \sim 20$. The bottom left panel shows the mass evolution in three temperature ranges: atomic gas with $10^{3.5} {\rm K} < {\rm T} < 10^{4.5} {\rm K}$ (purple), warm molecular-temperature gas with $10^{2} {\rm K} < {\rm T} < 10^{3.5} {\rm K}$ (yellow), and cold molecular-temperature gas with ${\rm T} < 10^{2} {\rm K}$ (green). All masses are normalized by the initial mass of the cloud. The grey vertical lines indicate important times in the mass evolution of the cloud. Corresponding to each grey line, we include a temperature slice from a simulation snapshot on the right to help visualize the morphology and composition of the cloud. The colors in the bottom left panel are also used in the four temperature slices on the right. In the first two temperature slices, when the cloud is not yet entrained, it contains mostly atomic gas. The molecular-temperature phase grows in the interior of the cloud after entrainment, as shown by the fourth temperature slice. A detailed analysis of this figure can be found in \autoref{sec:intermediate cloud}. A movie of this simulation can be found at \href{https://www.youtube.com/watch?v=rKtTNOZs7XA}{https://www.youtube.com/watch?v=rKtTNOZs7XA}.}
\label{fig:medium_cloud}
\end{figure*}

In this section, we analyze the evolution of a cloud with $r_{\rm initial} = 10 r_{\rm crit}$. Here, cloud material is defined to be all gas with $T < 10^{4.5}$K. 

In the top left panel of \autoref{fig:medium_cloud}, we see that the center of momentum velocity of the cloud (blue) asymptotically approaches the wind velocity (dashed black) as the cloud gets entrained. The \emph{average} density contrast (in brown; defined as $\chi = \left. \overline{\rho}_{\rm cloud} \right/ \rho_{\rm wind}$) never exceeds 100 before the cloud is entrained at $\left.t\right/ t_{\rm cc} \sim 20$. This is why the survival criterion for optically thin atomic clouds work well in our setup. After entrainment, the overdensity continues to rise as the molecular-temperature phase continues to grow. The bottom left panel of \autoref{fig:medium_cloud} shows the mass evolution in three different temperature ranges, with four grey vertical lines indicating important times in the mass evolution of the cloud. Corresponding to each grey line, we include a temperature slice from a simulation snapshot on the right to help visualize the morphology and composition of the cloud\footnote{We note that the simulation boxes we use are much longer and wider than what is shown in \autoref{fig:medium_cloud} such that cloud material does not flow out or interact with the boundaries. We are only showing a section of a temperature slice from a simulation snapshot that is the most representative of the morphology and composition of the cloud at the time of the snapshot.}. We see that the abundance of molecular-temperature gas is not monotonic. It grows rapidly at first, as one would expect from the short cooling time\footnote{Thus, although we started from a purely atomic cloud, our results are very similar to a more realistic situation where the initial cloud is partly atomic and partly molecular, as we have verified directly.}, but then is destroyed and only reforms later. How can we understand this? 

The initial radius of the cloud is large enough for its interior to self-shield ($N_{\rm initial}/N_{\rm crit} \sim 10$). Initially, the cloud is both shearing against the wind, and cooling in the optically thick interior. The rapid cooling in the optically thick interior explains the quick emergence of molecular-temperature gas (yellow and green curves) at the beginning of the cloud evolution. This initial molecular-temperature gas is concentrated at the head of the cloud, as shown by the temperature slice at $\left.t\right/ t_{\rm cc} = 6.43$ in \autoref{fig:medium_cloud}. As temperature drops by more than two orders of magnitude, molecular-temperature gas in this region contracts drastically to maintain pressure equilibrium. Thus, although the volume of the cold molecular-temperature phase appears to be tiny, it takes up a non-trivial fraction of the mass. 

As the cloud continues to evolve, it gradually develops a two-part morphology consisting of a head and a tail. The head, which contains the cold molecular-temperature nugget discussed above, is highly overdense and difficult to entrain. As a result, the outer atomic "shield" at the head is constantly sheared away and replenished by molecular-temperature gas which is mixing with hotter gas and heating up. Eventually, the head of the cloud becomes purely atomic, as shown in the temperature slice at $\left.t\right/ t_{\rm cc} = 12.22$. The tail of the cloud is more fragmentary and turbulent than the head, with even shorter mixing times and optically thin fragments. There is no stable region that can cool down to molecular temperatures, and the tail only consists of atomic gas. Thus, throughout the entrainment process ($t/t_{\rm cc}< 20$), the cloud {\it volume} is overwhelmingly dominated by $T \sim 10^4$K atomic gas, as evident in the temperature slices and consistent with the $\chi \sim 100$ average density contrast we observed above. This is true even when the {\it mass} of the molecular-temperature phase is non-negligible.

Once the cloud entrains, turbulence and mixing die down. Moreover, cloud fragments coagulate to reform into a single coherent structure with a cometary tail. Here, coagulation is driven by hot mass inflow into cold gas, which drives a gentle "wind" which attracts clouds to one another with an inverse-square force $F \propto A_1 A_2/r^2$, where the "monopole" is surface area \citep{Gronke23}. As the cloud grows in mass, the tail becomes optically thick and is able to cool down warm molecular temperatures (see temperature slice at $\left.t\right/ t_{\rm cc} = 19.94$), then to cold molecular temperatures(see temperature slice at $\left.t\right/ t_{\rm cc} = 48.08$) through self-shielding. Thus, after entrainment, the cloud overdensity rises continually. The entrained cloud at the end of the simulation has roughly comparable mass in the atomic and molecular-temperature phase.

\begin{figure*}
\centering
\includegraphics[width=\textwidth]{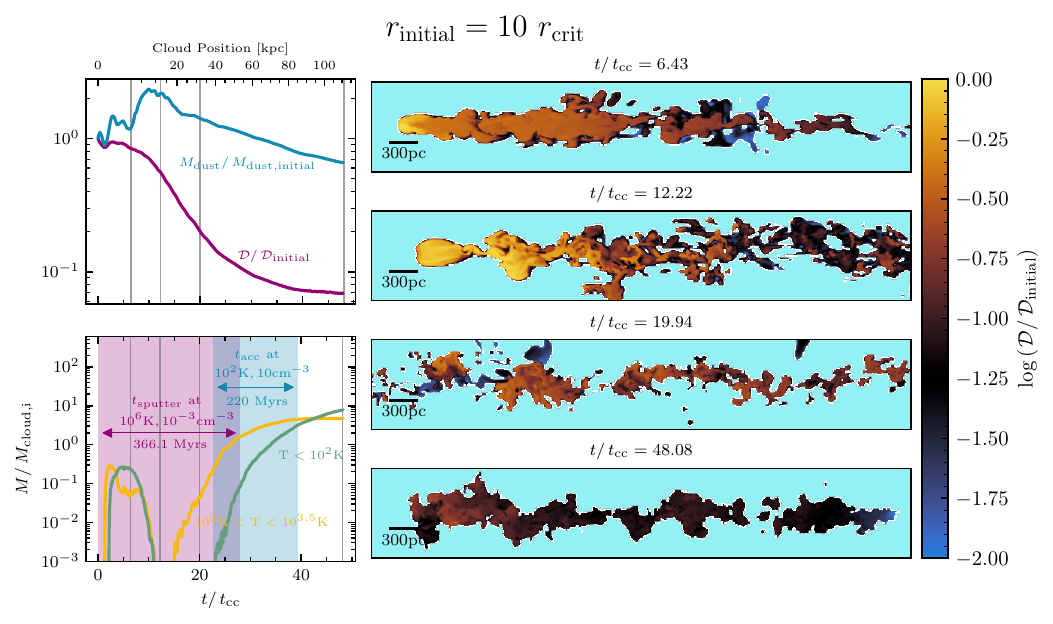}
\caption{Dust evolution analysis of a cloud with $r_{\rm initial} = 10 r_{\rm crit}$. We track dust evolution in our simulations using a passive scalar that accounts for thermal sputtering and mass accretion. On the right, we include simulation snapshots taken at the same times as the ones in \autoref{fig:medium_cloud}, but the (log-scaled) color map here indicates dust mass fraction of the cloud, normalized by the initial dust mass fraction. The wind, which contains no dust, is denoted by light blue. Each of the four simulation snapshots corresponds to a vertical grey line in the two left panels. The top left panel shows the average dust mass fraction (purple, normalized by the initial dust mass fraction) and dust mass (blue, normalized by the initial dust mass) evolution as a function of time (bottom horizontal axis) and distance travelled by the cloud (top horizontal axis). After normalization, both quantities are dimensionless and can thus be plotted in the same panel. The dimensionless form of dust mass fraction and dust mass allow us to better understand how these quantities evolve with respect to their initial values. The average dust mass fraction continuously declines as the cloud mixes with the wind during entrainment but stabilizes at $\left. \mathcal{D} \right/ \mathcal{D}_{\rm initial} \sim 0.07$ after $\left.t\right/ t_{\rm cc} \sim 40$, when dust accretion in the molecular-temperature phase of the entrained cloud is strong enough to compensate dust mass loss due to mixing. At the end of the simulation, the total dust mass has only slightly decreased to $\sim 0.7$ of its initial value. The bottom left panel shows the mass evolution in the warm molecular-temperature ($10^{2} {\rm K} < {\rm T} < 10^{3.5} {\rm K}$, yellow) and cold molecular-temperature (${\rm T} < 10^{2} {\rm K}$, green) phases, juxtaposed with a purple region whose width indicates the dust sputtering time in the hot wind (${\rm T} = 10^6 {\rm K}$, $n = 10^{-3} {\rm cm}^{-3}$) and a blue region whose width indicates the dust accretion time in the cold molecular-temperature phase (${\rm T} = 10^2 {\rm K}$, $n = 10 {\rm cm}^{-3}$). A detailed analysis of this figure can be found in \autoref{sec:intermediate cloud}.}
\label{fig:medium_cloud_dust}
\end{figure*}

The dust evolution is closely tied to the temperature evolution discussed above. The top left panel of \autoref{fig:medium_cloud_dust} shows the average dust mass (blue) and dust mass fraction (purple) of the cloud. The total dust mass is fairly constant and does not vary by more than a factor of two. By contrast, the dust mass fraction declines by more than an order of magnitude. This is largely due to dilution: while the dust mass is fairly constant, the cloud mass grows by a factor of $\sim 20$, by mixing with dust-free wind gas which subsequently cools. The minimal changes in the total dust mass suggest that both sputtering and grain growth are not very important. In the lower left panel of \autoref{fig:medium_cloud_dust}, we show the dust sputtering time (purple) in the hot wind (${\rm T} = 10^6 {\rm K}$, $n = 10^{-3} {\rm cm}^{-3}$), and dust accretion time (blue) in the cold molecular-temperature phase (${\rm T} = 10^2 {\rm K}$, ${\rm n} = 10 \, {\rm cm}^{-3}$), superposed against the mass evolution of the warm and cold molecular-temperature phases. These timescales are fairly long compared to cloud evolutionary timescales; at best, 2-3 e-folds are possible. In practice, their applicability is even more restricted. Sputtering is restricted to dust which mixes into hot gas, since $t_{\rm sputter} \propto T^{-5/2}$ is long at lower temperatures. However, by construction for a surviving cloud, most mixed gas cools, and there is a limited time $t_{\rm cool,mix} < t_{\rm sputter}$ for dust to sputter in mixed gas, which in any case is lower temperature than the wind. Grain growth requires high densities and mostly occurs in cold molecular-temperature gas ($T < 100$K). Thus it is only important once there {\it is} a significant cold, molecular-temperature component. It is responsible for the transient rise in dust mass (by factor of 2) at early times, which ceases once the molecular-temperature component is destroyed by mixing. Moreover, for grain growth to beat dilution, the dust accretion time $t_{\rm acc} \propto n^{-1} T^{-1/2} \propto P^{-1} T^{1/2}$ (where the last proportionality holds under isobaric conditions) must be shorter than the cloud growth time $t_{\rm cl,grow} \propto \chi r_{\rm cl}/v_{\rm mix}$ (where $v_{\rm mix} \sim c_{\rm s,c}$ is of order the cold gas sound speed; \citealt{gronke20-cloud}). This does not hold in this setup. Since $t_{\rm cl,grow}/t_{\rm acc} \propto \chi r_{\rm cl} P/T^{1/2}$, this suggests that for more overdense or larger clouds, or in higher pressure environments, grain growth can outcompete dilution. Even so, the present results already suggest that dust can be transported out to large circumgalactic distances without being destroyed by sputtering, since it is always cocooned by atomic gas. 

In the right column of \autoref{fig:medium_cloud_dust}, we plot four simulations snapshots taken at the same times as the ones in \autoref{fig:medium_cloud}, with the color map here indicating dust mass fraction (${\mathcal D}= \rho_{\rm dust}/\rho_{\rm gas}$) of the cloud, normalized by the initial dust mass fraction. The wind, which has no dust, is denoted by light blue. In the first simulation snapshot taken at $\left.t\right/ t_{\rm cc} = 6.43$, the cloud is just beginning to get deformed by the impinging wind. The dust mass fraction ${\mathcal D}/{\mathcal D}_{\rm initial} \sim 1$ at the head of the cloud and is lower in the newly formed tail, due to dilution. Up until $\left.t\right/ t_{\rm cc} = 12.22$, the head of the cloud maintains ${\mathcal D}/{\mathcal D}_{\rm initial} \sim 1$ because mixing with the wind is compensated by dust accretion in the molecular-temperature phase that formed and then got sheared away. At the same time, ${\mathcal D}/{\mathcal D}_{\rm initial}$  at the tail of the cloud starts to decline due to fragmentation and mixing with the wind. In the two final snapshots at $\left.t\right/ t_{\rm cc} = 19.94$ and 48.08, the cloud becomes entrained with a stable molecular-temperature interior, and dust becomes long-lived, with a fairly even distribution at ${\mathcal D}/{\mathcal D}_{\rm initial} \sim 0.1$ throughout the cloud. Although we do not continue to evolve this cloud for much longer, we expect this dust content to exist stably in the molecular-temperature phase of the cloud as it expands into the CGM.

\subsection{Clouds with different initial radii} \label{sec:Clouds with different initial radii}

In this section, we briefly discuss the evolution of clouds with $r_{\rm initial} = 3$ and $10 r_{\rm crit}$.

\begin{figure*}
\centering
\includegraphics[width=\textwidth]{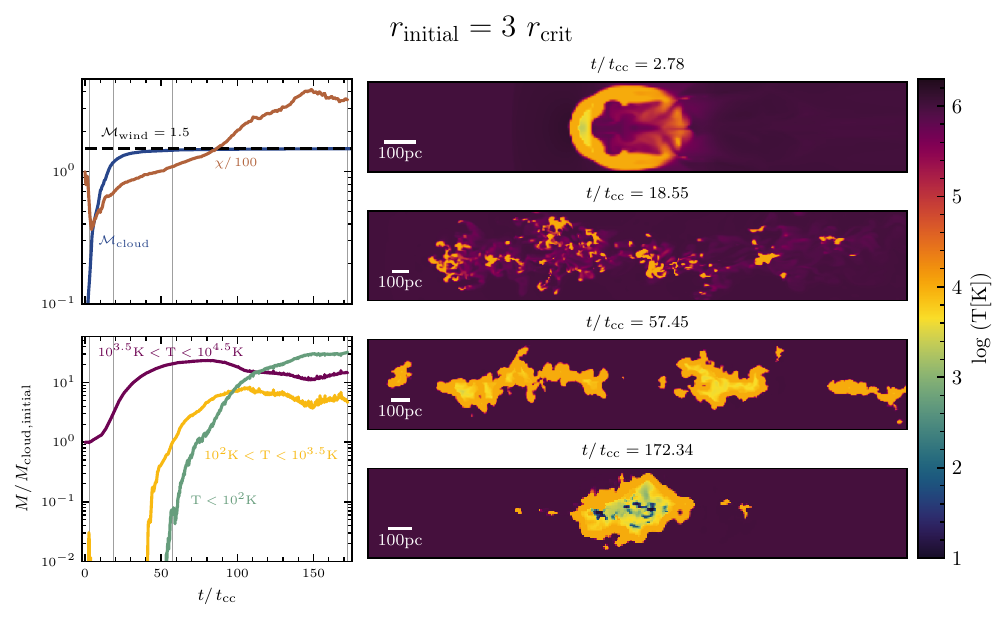}
\caption{Similar to \autoref{fig:medium_cloud}, but for a cloud with initial size $r_{\rm initial} = 3 r_{\rm crit}$. A movie of this simulation can be found at \href{https://www.youtube.com/watch?v=mooVSLa6S6w}{https://www.youtube.com/watch?v=mooVSLa6S6w}.}
\label{fig:small_cloud}
\end{figure*}

\begin{figure*}
\centering
\includegraphics[width=\textwidth]{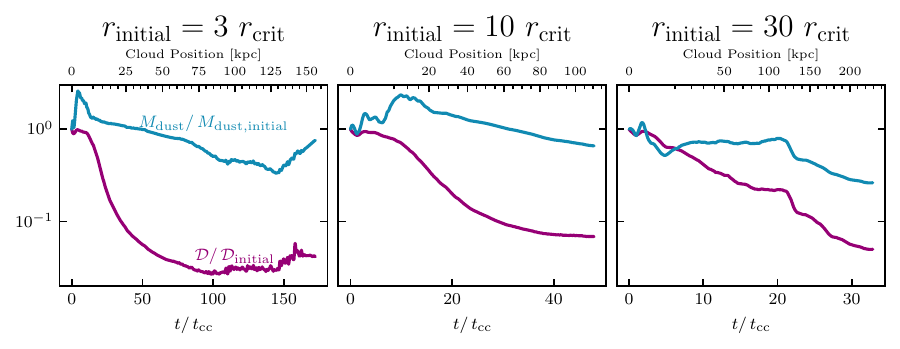}
\caption{Similar to the top left panel of \autoref{fig:medium_cloud_dust}, but for three clouds with initial size $r_{\rm initial} = 3$, 10, and 30 $r_{\rm crit}$. Note that since $t_{\rm cc}$ increases with cloud size, smaller clouds are run for longer in units of $t/t_{\rm cc}$, and are further along in their evolution. For all clouds, we expect dust content in entrained clouds to eventually grow, though this is not yet evident in the larger clouds.}
\label{fig:dust_evolution}
\end{figure*}

\autoref{fig:small_cloud} and the left panel of \autoref{fig:dust_evolution} show the evolution of a cloud with $r_{\rm initial} = 3 r_{\rm crit}$. Similar to the $r_{\rm initial} = 10 r_{\rm crit}$ case, this smaller cloud also maintains an average density contrast below 100 throughout cloud entrainment. The cloud shatters into many small fragments during entrainment. As the cloud gets entrained, the fragments re-coagulate and eventually become optically thick to develop a molecular-temperature interior. The entrained cloud experiences an order of magnitude growth in mass, and the atomic and molecular-temperature mass reach rough equipartition. Coagulation is particularly efficient for this smaller cloud, which has less of a cometary appearance.

As for dust evolution, this $r_{\rm initial} = 3 r_{\rm crit}$ cloud experiences a similar process of initial decline in dust mass fraction due to mixing with the wind and eventual stabilization of dust content in the molecular-temperature phase of the entrained cloud. $\left. \mathcal{D} \right/ \mathcal{D}_{\rm initial} \sim 0.04$ in the entrained cloud. Towards the end, there are hints of an increase in the dust mass fraction, indicating that grain growth is occurring faster than dilution. 

\begin{figure*}
\centering
\includegraphics[width=\textwidth]{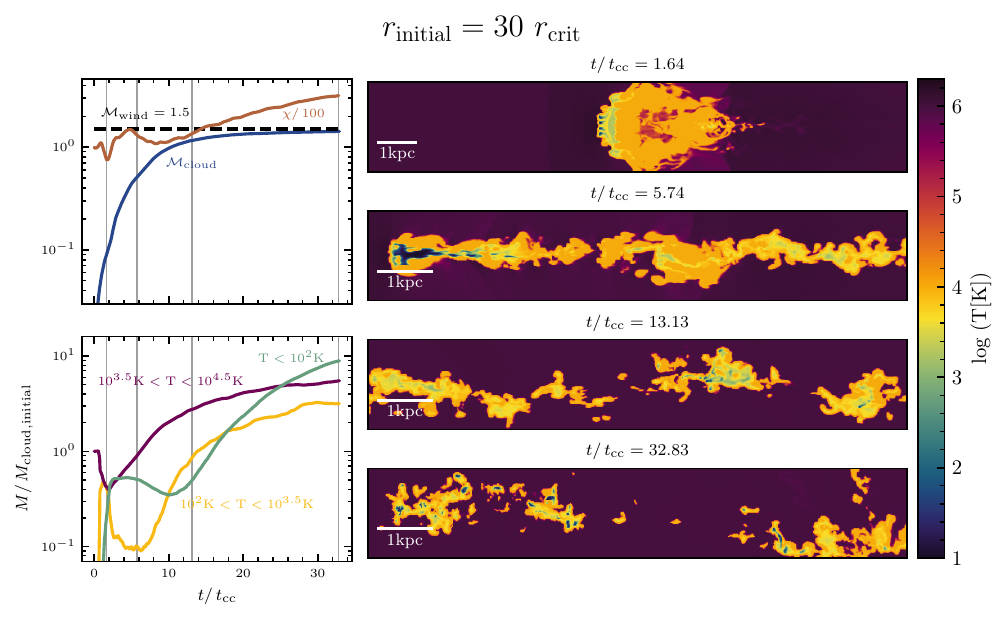}
\caption{Similar to \autoref{fig:small_cloud}, but for a cloud with initial size $r_{\rm initial} = 30 r_{\rm crit}$. A movie of this simulation can be found at \href{https://www.youtube.com/watch?v=rRdWv1xoLyk}{https://www.youtube.com/watch?v=rRdWv1xoLyk}.}
\label{fig:large_cloud}
\end{figure*}

\autoref{fig:large_cloud} shows the evolution of a cloud with $r_{\rm initial} = 30 r_{\rm crit}$. Like the $r_{\rm initial} = 10 r_{\rm crit}$ cloud, this cloud also develops a two-part morphology, with a head containing a molecular-temperature nugget that eventually gets sheared away, and a tail that grows by accreting atomic gas during entrainment. Although the average density contrast exceeds 100 slightly at $\left.t\right/ t_{\rm cc} \sim 15$, the cloud velocity evolution is not significantly affected because by the time the density contrast exceeds 100, the velocity difference between the cloud and the wind is already significantly reduced. However, for this cloud, the tail is able to maintain a stable, self-shielded interior before the cold nugget at the head is depleted, and we see molecular-temperature gas in both the head and the tail in the temperature slice at $\left.t\right/ t_{\rm cc} = 5.74$ in \autoref{fig:large_cloud}. This means, unlike the $r_{\rm iniial} = 10 r_{\rm crit}$ cloud, molecular-temperature gas is never depleted in this $r_{\rm initial} = 30 r_{\rm crit}$ cloud. At the end of the simulation, atomic and molecular-temperature gas again reach rough equipartition in mass. 

The dust evolution is similar to the previously discussed cases: dust accretion in the molecular-temperature phase at the head of the cloud compensates for the dust lost due to mixing until the head of the cloud is destroyed. Although the dust mass is stable from $t/t_{\rm cc}$, at $t/t_{\rm cc} \sim 20$, there is a significant decrease in dust mass. This is coincident with a change in cloud morphology, where the cloud appears to break up into smaller pieces. Overall, however, these variations in dust content are not large. At the end of the simulation, the entrained cloud has $\sim 0.05$ times the initial dust mass fraction and $\sim 0.5$ the initial dust mass, similar to previous cases.  

In conclusion, all clouds we simulated using our fiducial setup maintain a low overdensity during cloud entrainment that facilitates survival and entrainment. Gas is able to cool to cold molecular temperatures in the self-shielded interior of the entrained clouds. During the entrainment process, the amount of cool gas present does depend on cloud size. The $r_{\rm initial} = 3 r_{\rm crit}$ cloud essentially has no $T < 10^{3.5} {\rm K}$ gas until it is completely entrained; the $r_{\rm initial} = 10 r_{\rm crit}$ cloud forms $T < 10^{3.5} {\rm K}$ gas at the head, gets destroyed, and eventually reforms $T < 10^{3.5} {\rm K}$ gas at the tail; the $r_{\rm initial} = 30 r_{\rm crit}$ cloud always hosts $T < 10^{3.5} {\rm K}$ gas because the dust  destruction at the head happens after the dust formation at the tail. Dust is protected from thermal sputtering and the total dust mass of the entrained cloud is generally $\sim 50\%$ of the initial dust mass, although the mass fraction is diluted to $\mathcal{D}/\mathcal{D}_{\rm initial} \sim 5\%$ as hot, dust free wind material cools onto the cloud. Note that from the slice plots and the dust evolution figure, it may appear that largest cloud is more fragmentary, and the dust abundance has not quite equilibrated, and appears slowly declining. These two effects are related, since more fragmentary clouds mix more with the hot medium and lose more dust.  However, we caution that the simulations are run for different $t/t_{\rm cc}$, since $t_{\rm cc} \propto r$ is longer for larger clouds. If we compare the entrained clouds at comparable values of $t/t_{\rm cc}$, their morphology and dust evolution appear broadly comparable. If we run the larger cloud simulations for longer, we expect the fragments in the larger cloud cases to also coagulate, and for the dust content to stabilize and grow. 

\section{Understanding Molecular-Temperature Cloud Entrainment} \label{sec:understading molecular cloud entrainment}

In this section, we explain the abundance of atomic gas seen during cloud entrainment, which is crucial to reducing the overdensity of the cloud and thereby enabling survival and entrainment. 

\subsection{Quantifying Cloud Composition Using Temperature PDFs} \label{sec:cloud composition}

The most important observation we can make about the clouds in our simulations is that they are almost purely atomic during entrainment. This is evident from the cloud overdensity evolution (e.g. see top left panel of \autoref{fig:medium_cloud}) and the temperature slices at important times shown in \autoref{fig:medium_cloud}. In this section, we further quantify this by analyzing cloud composition through temperature PDFs.

\begin{figure*}
\centering
\includegraphics[width=\textwidth]{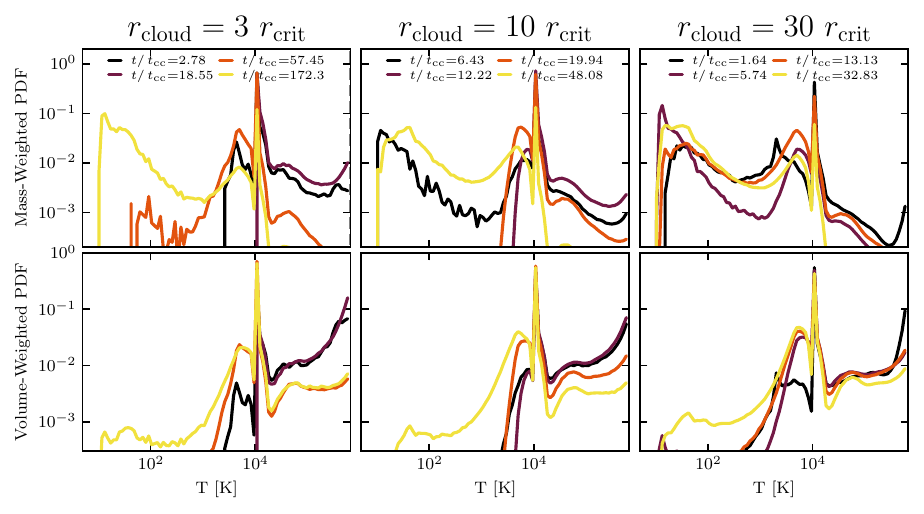}
\caption{Mass- and Volume-weighted probability density function (PDF) of the temperature distribution of the three clouds we analyzed before. All histograms are normalized. For each cloud, we consider 4 evolution stages that correspond to the times when the snapshots in the previous figures are taken. Lighter color indicates later times. The temperature PDFs are skewed to atomic temperatures at early times and only occupy the full dynamical range in temperature down to 10K at late times once the clouds are fully entrained. At these late times, the shape of the mass-weighted temperature PDF resembles the shape of the isobaric cooling time profile shown in \autoref{fig:cooling_curve}. We comment on the late time temperature PDFs in \autoref{sec:The Damkohler Number and Atomic Gas Production Through Mixing}. Note that the volume-weighted PDFs heavily favor atomic gas, even at late times when the clouds are entrained and can cool down to molecular temperatures.}
\label{fig:volume_and_mass_histograms}
\end{figure*}

In \autoref{fig:volume_and_mass_histograms}, we plot the mass- and volume-weighted probability density function of the temperature distribution (henceforth referred to as the temperature PDF) of the three clouds we analyzed above, with the different colors indicating different times in the simulations. For our fiducial cloud with $r_{\rm initial} = 10 r_{\rm crit}$ (middle column of \autoref{fig:volume_and_mass_histograms}), we can see that despite some initial cooling that populates the molecular temperatures at $\left. t \right/ t_{\rm cc} =6.43$ (black), the entire cloud soon becomes atomic after there is sufficient time for the wind to deform and elongate it. This is shown by the temperature PDF at  $\left. t \right/ t_{\rm cc} =12.22$ (purple), which truncates at $\sim 5000$K. The cloud does not occupy the full dynamic range of temperatures down to 10K until long after entrainment at $\left. t \right/ t_{\rm cc} =48.08$ (yellow). 

The two other clouds with different initial radii shows a similar behavior. The $r_{\rm initial} = 3 r_{\rm crit}$ does not have time to cool down to molecular temperatures before the wind shears it and turns it to be purely atomic. We see growth of the molecular-temperature phase starting from $\left. t \right/ t_{\rm cc} =57.45$ (orange), again long after entrainment. As for the $r_{\rm initial} = 30 r_{\rm crit}$ cloud, the temperature PDFs can be deceptive because it does not distinguish between molecular-temperature gas that forms at different parts of the cloud. The molecular-temperature phase we see at early times in the temperature PDFs reside in the head of the cloud, and it gets sheared away as the more overdense cloud head takes longer to entrain. The remnant of the head that actually survives and gets entrained is again atomic. The molecular-temperature phase at late times is then formed in the interior of the entrained cloud.

We briefly note that at late times in the entrained clouds, the mass-weighted temperature PDFs has gas throughout the molecular temperature range (10K to $\sim$ 5000K) instead of only having gas at the temperature floor of 10K. Given the fact that cloud cooling times are much shorter than these evolutionary timescales ($\tau_{\rm cool} \lsim t_{\rm cc}$), this requires explanation. Furthermore, the shape of the late time temperature PDFs tracks features in the shape of the isobaric cooling time profile shown in \autoref{fig:cooling_curve}. The bump in the PDF at $\sim 5000$K corresponds to the cooling bottleneck discussed extensively above, and similarly, at $\sim 45$K we simultaneously see another bump in the PDF and a steep increase in the cooling time. More importantly, the volume-weighted PDFs heavily favor atomic gas, even at late times when the clouds are entrained and can cool down to molecular temperatures. This is the key reason why cloud overdensities remain small. Thus, even when atomic and molecular-temperature components are comparable by mass, atomic gas still heavily dominates by volume, and the cloud overdensity is still not much greater than 100.

Multiple papers studying the interaction between turbulence and radiative cooling in the ISM have reported temperature PDFs very similar to ours. Perhaps the most directly comparable is \cite{Saury:2014}, who studied the phase transition from the warm neutral medium to the cold neutral medium (i.e., gas in the temperature range $10 \, {\rm K} < T < 10^{4} \, {\rm K}$) under the influence of thermal instability and a mixture of compressive and solenoidal turbulent forcing, with $v_{\rm turb} \sim 10 \, {\rm km \, s^{-1}}$ in $(1024)^3$ simulations. Figure 5 and 14 in their paper shows the temperature PDFs obtained from their simulations, which have a broad distribution of gas at all temperatures between 10K and $10^4$K, similar to the end state of our entrained clouds. Their mass-weighted PDFs exhibit an accumulation of gas at $\sim 45$K and $\sim 5000$K just like ours, and the maximum and minimum in the PDFs differ by less than an order of magnitude, indicative of the large amount of intermediate temperature gas. In earlier work, \cite{Audit:2005} also noted the large fraction of gas in the thermally unstable regime in simulations of driven turbulence . The left column of their Figure 8 is directly comparable to our temperature PDFs. As they increase the amplitude of turbulent driving, the amount of cold gas is increasing suppressed (in particular, the second peak at the temperature floor), as we also see during the entrainment stage. Indeed, in reviewing the broad temperature PDFs seen in simulations, \cite{vázquez-semadeni:2012} argued that both turbulent shock compression and mixing are so effective in populating classically forbidden (thermally unstable) temperature ranges, that the notion of discrete phases in a highly turbulent medium is not appropriate; instead, the word "phase" should simply be used to denote a certain temperature or density range.

In conclusion, a unifying theme of cloud evolution that is confirmed by the shape of the temperature PDFs is that the clouds are mostly atomic with a low overdensity of $\sim$ 100 during entrainment. The molecular-temperature phase is formed later, only after the cloud is entrained. This two-stage evolutionary history clears facilitates molecular-temperature gas entrainment, but the real question is: why is there so much atomic gas during cloud entrainment? What is providing this "lifeline" for molecular-temperature gas?

\subsection{What Roles Do Photoionization \& the Cooling Bottleneck Play?} \label{sec:What Role Does Photo-ionization and the Cooling Bottleneck Play?}

It seems natural to attribute the abundance of atomic gas to either photoionization or features in the isobaric cooling time profile. Photoionization produces an optically thin atomic skin for the clouds, which reduces the cloud overdensity as desired. Furthermore, as we pointed out in \autoref{sec:radiative cooling}, the isobaric cooling time profile rises sharply just below $10^4$K due to hydrogen recombination (see \autoref{fig:cooling_curve} for more details). This cooling bottleneck prolongs the cooling process down to molecular temperatures and leads to an accumulation of gas at $\sim 5000$K. Naively, these factors seem important in producing large amounts of atomic gas at $\sim 10^4$K and ensuring cloud survival.

To fully understand the roles of photoionization and the cooling bottleneck, we take our $r_{\rm initial} = 10 r_{\rm crit}$ cloud analyzed in \autoref{sec:intermediate cloud} as a point of comparison and run three more simulations with all else held equal but removing photoionization, removing the cooling bottleneck, and removing both. Removing photoionization is straightforward: we simply allow all cloud cells in our simulations to cool down to the temperature floor at 10K, without distinguishing between optically thin and optically thick regions. Removing the cooling bottleneck at $\sim 5000$K, on the other hand, is a bit more tricky because it requires a modification to the standard cooling time profile. Specifically, we modify the cooling time profile such that the cooling times between $\sim 200$K and $\sim 2 \times 10^4$K is smoothly connected by a power-law. We illustrate this change graphically in \autoref{fig:cooling_curve} using the dashed magenta line labeled "No Cooling Bottleneck". 

\begin{figure*}
\centering
\includegraphics[width=\textwidth]{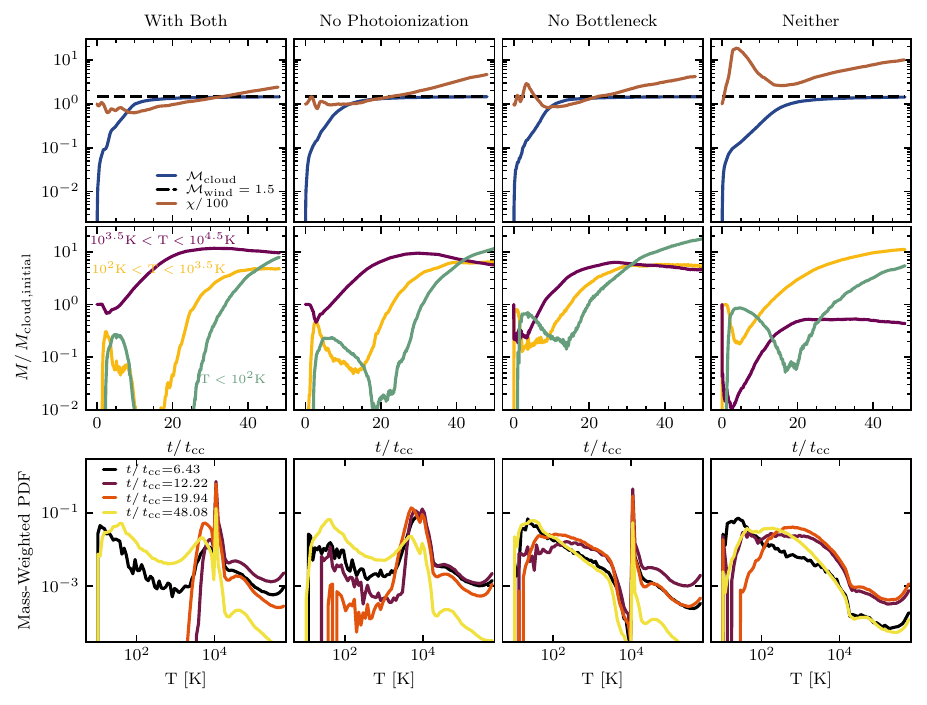}
\caption{We examine the effects of photoionization and the cooling bottleneck by cross-comparing simulations with or without them. The leftmost column is obtained from our fiducial simulation with $r_{\rm initial} = 10 r_{\rm crit}$ (analyzed in \autoref{sec:intermediate cloud} and \autoref{fig:medium_cloud}). The three columns to the right are obtained from simulations with all else held equal but without photoionization, without the cooling bottleneck, and without both, respectively. The top row shows the cloud velocity and overdensity evolution, the middle row shows the mass evolution in three temperature bins, and the bottom row shows temperature PDFs for the cloud at important times. All four clouds get entrained in a comparable timescale, experience an order of magnitude mass growth, and exhibit similar temperature PDFs after entrainment. These results tell us that although photoionization and the cooling bottleneck both promote the dominance of atomic gas during entrainment, neither is a necessary condition for cloud survival and entrainment. Molecular-temperature gas can survive and get entrained even when neither is present.}
\label{fig:self_shielding_and_cooling_bottleneck}
\end{figure*}

We examine the effects of photoionization and the cooling bottleneck by cross-comparing simulations with or without them. In \autoref{fig:self_shielding_and_cooling_bottleneck}, we plot the evolution of cloud velocity, overdensity, and mass as a function of time as well as temperature PDFs for each of the simulations we ran (fiducial, no photoionization, no cooling bottleneck, neither). Surprisingly, all four clouds get entrained on a comparable timescale and experiences an order of magnitude mass growth after entrainment. The average density contrast of the cloud is only $\chi \sim$ 100 in the first three columns where we have at least either photoionization or the cooling bottleneck. Even when neither is present, the average density contrast only briefly spikes to $\sim 1000$ before falling back down to slightly over 100 during cloud entrainment. This delays but does not prevent entrainment. The temperature PDFs at late times for the entrained clouds also look similar. They are flat rather than dominated by molecular temperature gas.

Based on these results, it is fair to conclude that photoionization and the cooling bottleneck both promote the dominance of atomic gas during entrainment. When either is present, the cloud is mostly atomic during the entrainment process. By contrast, when both are removed, the cloud mostly consists of molecular-temperature gas. In reality, of course, {\it both} will always be present; clouds during the entrainment process should be predominantly atomic and therefore easily entrained. While the role of photoionization is easily understood, the importance of the cooling bottleneck (which, based on the second row of \autoref{fig:self_shielding_and_cooling_bottleneck}, has a greater influence on the mass evolution of the various phases of the cloud) is harder to understand. The cooling time of $\sim 10$ Myr it imposes in our setup is at most of order a cloud-crushing time, much shorter than the simulation run-time. It seems odd that it should prevent gas from cooling down to molecular temperatures. Furthermore, the experiment of removing both photoionization and the cooling bottleneck shows that a predominantly atomic cloud is a sufficient but not necessary condition for cloud survival and entrainment. The molecular-temperature cloud during entrainment is still fairly warm (with a median temperatures of $T \sim 1000$K and thus overdensities of $\chi \sim 1000$, not $\chi \sim 10^4 - 10^5$). We now try to understand these results. 

\subsection{The \Da Number and Atomic Gas Production Through Mixing} \label{sec:The Damkohler Number and Atomic Gas Production Through Mixing}

The tendency to cool down to a single-phase stable medium at the temperature floor versus maintaining large amounts of intermediate (atomic) temperature gas often boils down to a competition between mixing and cooling. A glance at \autoref{fig:cooling_curve} shows that isobarically cooling from $10^4$K to 10K only takes less than 10 Myrs, which is just a few cloud-crushing times for even the smallest cloud we simulate. Thus, maintaining the cloud to be atomic throughout entrainment is a non-trivial task. 

To understand whether the cloud is constantly "stirred" during entrainment to overcome its tendency of rapid cooling, we introduce a useful dimensionless parameter that characterizes the competition between mixing and cooling: the Damk$\ddot{\rm o}$hler number (Da) \citep{Damköhler:1940}. The Damk$\ddot{\rm o}$hler number is commonly used in the turbulent combustion literature and was applied to turbulent mixing layers in \cite{tan21}. It is defined as

\begin{align}
    {\rm Da} = \frac{t_{\rm mix}}{\tau_{\rm cool}} = \frac{\left. L \right/ v_{\rm turb}}{\tau_{\rm cool}}, \label{eq:damkohler number}
\end{align}

where L is some characteristic outer lengthscale used to calculate the eddy turnover time or mixing time $t_{\rm mix}$, $v_{\rm turb}$ is the turbulent velocity in the cloud, and $\tau_{\rm cool}$ is the characteristic cooling time of the medium given by equation \ref{eq:tau_cool}. The Damk$\ddot{\rm o}$hler number separates out two regimes. When Da $>$ 1, material cools faster than it mixes, and the medium can settle into the thermally stable phases. In the temperature ranges we are interested in, we find Da $>$ 1 gas to be thermally bi-stable, with both atomic ($\sim 5000$K) and molecular-temperature ($\sim 10$K) phases. When Da $<$ 1, mixing is sufficiently strong that cooling down to the thermally stable phases is suppressed, and instead there is an abundance of intermediate temperature gas. In our case, this intermediate temperature gas happens to be at atomic temperatures ($\sim 10^4$K).

To determine which regime our cloud is in, we calculate the Damk$\ddot{\rm o}$hler number in our fiducial $r_{\rm initial} = 10 r_{\rm crit}$ cloud at $\left. t \right/ t_{\rm cc} = 12.22$ (this corresponds to the second temperature slice in \autoref{fig:medium_cloud}). We emphasize that at the time we choose to evaluate the Damk$\ddot{\rm o}$hler number, the cloud is still undergoing entrainment, with a significant relative velocity ($\sim 0.5 c_{\rm s,hot}$) with respect to the wind. The cloud is mostly atomic during (not after) entrainment, and this evolutionary stage controls the eventual survival of molecular-temperature gas in the entrained cloud. 

Calculating the Damk$\ddot{\rm o}$hler number using \autoref{eq:damkohler number} requires appropriate choices of L, $v_{\rm turb}$, and $\tau_{\rm cool}$. For L, the characteristic outer lengthscale, we choose the transverse dimension of the head of the cloud, as annotated in white on the top panel of \autoref{fig:Damkohler_number_medium_cloud_during_entrainment}. Note that this choice of L is similar to the initial radius of the cloud at 300pc. Of course, this is a crude treatment that leads to overestimation of the Damk$\ddot{\rm o}$hler number at the tail of the cloud because the tail is clearly more fragmented than the head. However, since we will argue that mixing is important, this is a conservative choice since at worst it underestimates the strength of mixing.

\begin{figure}
\centering
\includegraphics[width=\columnwidth]{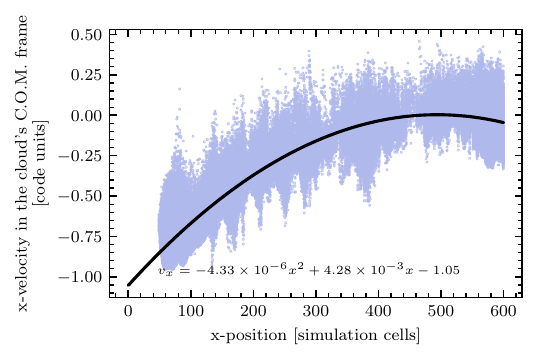}
\caption{Scatter plot of $v_{\rm x}$ as a function of $x$. Each purple dot represents a cloud cell in our simulation snapshot. Note how $\langle v_r \rangle$ depends on $x$. This is because the head of cloud once contained a dense molecular-temperature core and thus experienced a much smaller acceleration due to the wind at the beginning of the simulation. The best-fit function is a parabola, which we plot in black with the analytic expression annotated.}
\label{fig:vx_vs_x_position_medium_cloud_during_entrainment}
\end{figure}

\begin{figure*}
\centering
\includegraphics[width=\textwidth]{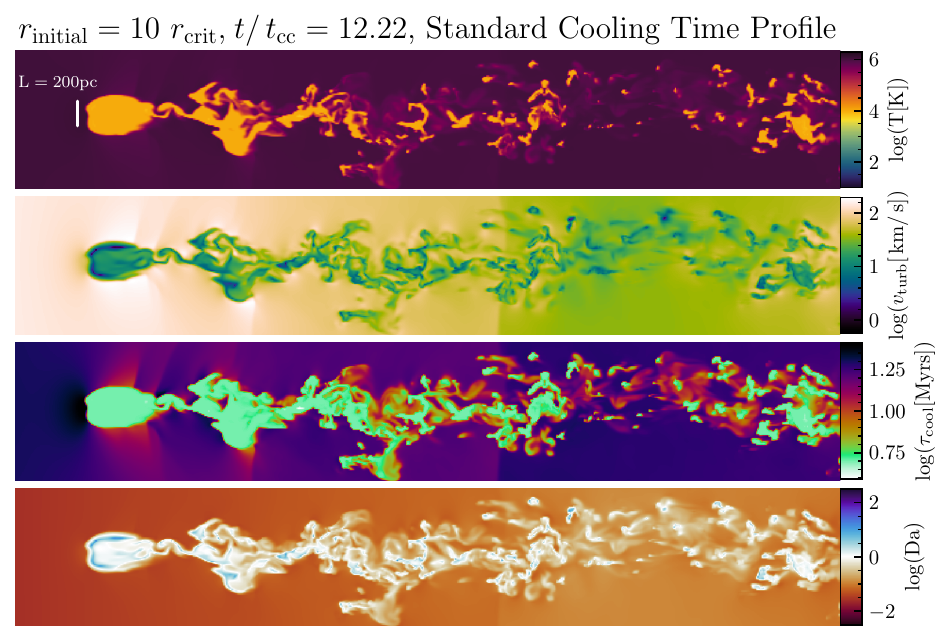}
\caption{Damk$\ddot{\rm o}$hler number calculation for the $r_{\rm initial} = 10 r_{\rm crit}$ cloud during cloud entrainment ($\left. t \right/ t_{\rm cc} = 12.22$). The first panel shows a temperature slice of the cloud, with L, the characteristic outer lengthscale used to calculate the mixing time, annotated in white. We choose L=200pc, which is intended to be roughly the transverse dimension of the cloud. The turbulent velocity ($v_{\rm turb}$) in the cloud is plotted in the second panel. Each component of $v_{\rm turb}$ is obtained by subtracting out the bulk laminar motion of the cloud. We note that in the direction parallel to the wind, the bulk motion of the cloud varies with position in the cloud. We fit this relationship using a parabola (see \autoref{fig:vx_vs_x_position_medium_cloud_during_entrainment} for more details) so that we subtract out the appropriate bulk velocity for different parts of the cloud. The third panel shows the time needed to cool to molecular temperatures ($T\sim 45$K), which we call $\tau_{\rm cool}$. This timescale is sensitive to the shape of the cooling time profile in the entire temperature range through which the gas (hypothetically) cools through and is distinct from the isobaric cooling time $t_{\rm cool}$. Finally, we plot the Damk$\ddot{\rm o}$hler number in the fourth panel. Most of the cloud has Da$<1$, which means mixing wins over cooling to keep the cloud atomic throughout entrainment. This is crucial to cloud survival and molecular-temperature gas formation in the entrained cloud.}
\label{fig:Damkohler_number_medium_cloud_during_entrainment}
\end{figure*}

To obtain the turbulent velocity $v_{\rm turb}$, we must first remove bulk velocity components. We first consider the direction parallel to the wind, which we define to be the x-direction. It is essential to realize that at this evolutionary stage, different parts of the cloud are moving at different velocities in the x-direction. The head of the cloud, which initially contained dense molecular-temperature gas (see first temperature slice of \autoref{fig:medium_cloud}), has been sitting like a large overdensity rock, and only recently transformed to atomic gas. It has a large velocity difference with respect to the wind. On the other hand, the tail of the cloud is atomic throughout its evolution and is already significantly entrained. The bulk velocity gradients across the cloud are larger than in  well-studied optically thin cloud-crushing simulations. Due to this complication, standard approaches of finding $v_{\rm turb}$, such as performing Gaussian filtering on the entire cloud, computing the velocity structure function \citep{Abruzzo:2022b}, or simply subtracting out the same bulk velocity for all parts of the cloud do not work. Instead, we fit the cloud's bulk velocity in the $x$-direction $\langle v_{\rm x}(x)\rangle$ as a function of x and subtract it off. This ensures that the appropriate bulk velocity is subtracted out at different parts of the cloud. \autoref{fig:vx_vs_x_position_medium_cloud_during_entrainment} shows our fit to the bulk velocity in the $x$-direction. From this, we obtain: 
\begin{equation}
    v_{\rm turb, x}(x) = v_{\rm x}(x) - v_{\rm bulk, x}(x).
\end{equation}
By contrast, the bulk velocity in the $y,z$ directions has relatively little positional dependence \footnote{Given the complications with the bulk velocity in the x-direction, it is tempting to only use the transverse components in the $v_{\rm turb}$ calculation. However, we note that there is some anisotropy in $v_{\rm turb}$, with the x-component a bit larger than the other components because it is parallel to the impinging wind. For the cloud shown in \autoref{fig:Damkohler_number_medium_cloud_during_entrainment}, we have found $\langle v_{\rm turb,x} \rangle \sim 11 \left. {\rm km} \right/ {\rm s}$ while $\langle v_{\rm turb,y} \rangle \sim \langle v_{\rm turb,z} \rangle \sim 7.5 \left. {\rm km} \right/ {\rm s}$. Thus, we include the x-component in our $v_{\rm turb}$ calculation to fully reflect turbulent velocities.}; we can ignore it to obtain:
\begin{align}
    v_{\rm turb, y} = v_{\rm y} - v_{\rm bulk, y},\\
    v_{\rm turb, z} = v_{\rm z} - v_{\rm bulk, z}.
\end{align}
Finally, $v_{\rm turb}$ can then be obtained by
\begin{align}
    v_{\rm turb} = \sqrt{v_{\rm turb, x}^2 + v_{\rm turb, y}^2 + v_{\rm turb, z}^2}
\end{align}

Now we need to compute $\tau_{\rm cool}$. In standard cloud crushing simulations, the timescale for a cell at temperature $T$ to cool down to the floor temperature is given by the its isobaric cooling time $t_{\rm cool}(T)$. This is because for $T> 10^4$K, the isobaric cooling time $t_{\rm cool}(T)$ declines monotonically as temperature falls. Thus, the rate-limiting step is the cooling time at the current (highest) temperature of the cell. This is similar to how the timescale for mixing in Kolmogorov turbulence is given by the outer eddy turnover time  $t_{\rm turb} \sim L/v \propto L^{2/3}$, since $t_{\rm turb}$ declines monotonically with scale. However, once $t_{\rm cool}(T)$ no longer decreases monotonically at lower temperatures, the instantaneous cooling time is no longer a good estimate of the time required to reach the floor temperature.  For instance, consider a cloud cell at $10^4$K. Reading the standard $t_{\rm cool}$ profile on the top panel of \autoref{fig:cooling_curve} tells us that its isobaric cooling time $t_{\rm cool}$ is $\sim 10^{-2}$ Myrs. However, $\tau_{\rm cool}$ here is much longer because as the $10^4$K cloud cell cools, it encounters the cooling bottleneck at $\sim 5000$K, which is nearly three orders of magnitude higher than $t_{\rm cool}$ at $10^4$K. In other words, the process of cooling down to molecular temperatures is sensitive to the shape of the cooling time profile in the entire temperature range in which the cloud cools through, which makes $t_{\rm cool}$ a bad proxy of the characteristic cooling time. Instead, we take the characteristic cooling time to be $\tau_{\rm cool}$, the time needed to cool down to the minimum of $t_{\rm cool}$ in the molecular temperature range at $T = 45$K. $\tau_{\rm cool}$ can be obtained from $t_{\rm cool}$ using \autoref{eq:tau_cool}. We plot $\tau_{\rm cool}$ profiles in the bottom panel of \autoref{fig:cooling_curve}. Note that the $\tau_{\rm cool}$ profile accounts for the suppression of cooling to temperatures below $\sim 5000$K by the cooling bottleneck, which is the behavior we want to capture.\footnote{The cooling bottleneck is similar to the maximum cooling time $t_{\rm cool,max}$ which \citet{farber22} identify in their analysis.}

We then calculate the Damk$\ddot{\rm o}$hler number for our fiducial $r_{\rm initial} = 10 r_{\rm crit}$ cloud during entrainment using \autoref{eq:damkohler number}. It is plotted in the bottom panel of \autoref{fig:Damkohler_number_medium_cloud_during_entrainment}. Most of the cloud, colored in white and light brown, has ${\rm Da}<1$. This means the mixing time is shorter than the cooling time, and vigorous mixing prevents the cloud from cooling down to molecular temperatures, keeping the cloud atomic during entrainment. Turbulent dissipation can also play an important role in heating the gas and maintaining it above $T \sim 5000$K. Note that the turbulent velocities are transonic to mildly supersonic, and thus the turbulent dissipation rate $\epsilon \sim \rho v^2/t_{\rm turb} \gsim \rho c_{\rm s}^2/t_{\rm cool} \sim P_{\rm th}/\tau_{\rm cool}$, i.e. the turbulent heating rate is comparable to the radiative cooling rate\footnote{We have used $\mathcal{M} \gsim 1$, i.e. $v \gsim c_{\rm s}$, and Da$\lsim 1$, i.e. $t_{\rm turb} \lsim t_{\rm cool}$.}. Turbulent mixing and dissipation have simplified the molecular-temperature gas entrainment problem to be essentially identical to the well-studied optically thin atomic cloud entrainment scenario, which is well known to survive and entrain once $r > r_{\rm crit}$. 

\begin{figure*}
\centering
\includegraphics[width=\textwidth]{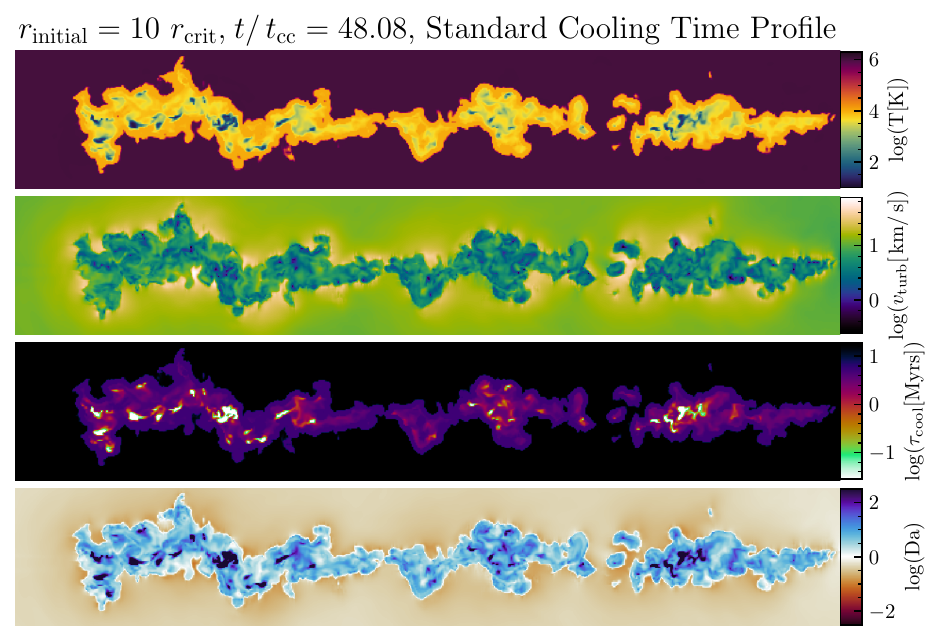}
\caption{Similar to \autoref{fig:Damkohler_number_medium_cloud_during_entrainment}, but at $\left. t \right/ t_{\rm cc} = 48.08$, when the cloud is entrained. Through most of the cloud, cooling wins over mixing because $\tau_cool$ is comparable to that during entrainment while $v_{\rm turb}$ dropped significantly after the shear of the wind falls. Da $\sim$ 10 in this entrained cloud, and it develops a multi-phase composition.  We suspect that this cloud composition represents a saturated state of thermal instability in a multi-phase turbulent medium in the temperature and \Da number range we are considering, driven by the shape of the cooling curve. }
\label{fig:Damkohler_number_medium_cloud_entrained}
\end{figure*}

What happens after entrainment? In \autoref{fig:Damkohler_number_medium_cloud_entrained}, we plot temperature, $v_{\rm turb}$, $\tau_{\rm cool}$, and the Damk$\ddot{\rm o}$hler number for our fiducial $r_{\rm initial} = 10 r_{\rm crit}$ cloud after entrainment. We see that through most of the cloud, cooling wins over mixing because $\tau_{\rm cool}$ is comparable to that during entrainment while $v_{\rm turb}$ dropped significantly after the shear of the wind falls. As a result, Da $\sim 10$ in the cloud, and the cloud develops a multi-phase composition. We suspect that this multi-phase cloud composition represents a saturated state of thermal instability in a multi-phase turbulent medium in the temperature and Damk$\ddot{\rm o}$hler number range we are considering, driven by the shape of the cooling curve.
We elaborate on this further below.
  
\begin{figure}
\centering
\includegraphics[width=\columnwidth]{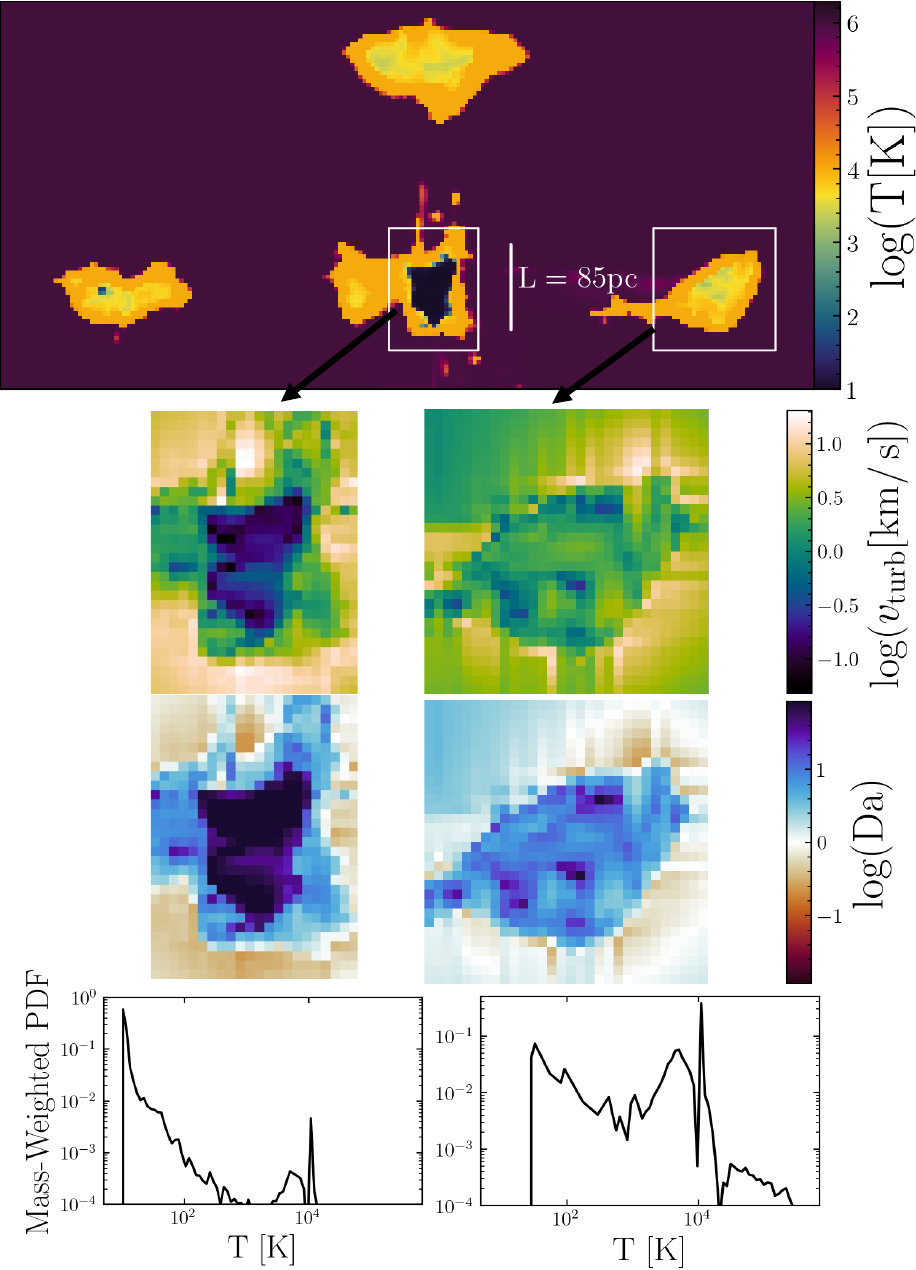}
\caption{Snapshot of our static cloud simulation. The static cloud is initialized at $10^2$K in pressure equilibrium with the hot background. At the beginning of the simulation, the cloud breaks up into a central piece that contains most of the mass and several satellites. This snapshot shows these pieces after they have interacted and stabilized in configuration. Below the simulation snapshot, we plot $v_{\rm turb}$, the Damk$\ddot{\rm o}$hler number, and the temperature PDF for the central cloud and one of the satellites. The central cloud is basically stationary and has a low $v_{\rm turb}$, while the satellite's $v_{\rm turb}$ is an order of magnitude higher. As a result, the \Da number of central cloud is an order of magnitude above the satellite cloud. This difference is reflected in their phase structure. The central cloud is singled-phased at the temperature floor, but the satellite cloud maintains a multi-phase composition despite having plenty of time to cool. The \Da number, composition, and temperature PDF of the satellite cloud resembles that of the entrained cloud in our fiducial simulation.}
\label{fig:static_cloud}
\end{figure}

Even after entrainment, the cold molecular-temperature gas fraction rises on a relatively slow timescale (many $t_{\rm cc}$), rather than on $\tau_{\rm cool} \lsim t_{\rm cc}$. This could be due to residual decaying turbulence from the entrainment process. In fact, the process of coagulation itself can provide enough gas motions and mixing to delay the appearance of gas at the temperature floor. We can mimic an fully entrained cloud with no relative shear or hot gas motions by simulating a static cloud in a uniform hot $T \sim 10^6$K background. We initialize the static cloud at $10^2$K, in pressure equilibrium with the hot background, and seed percent-level fluctuations. At the beginning of the simulation, the static cloud cools, falls out of pressure balance (since the cooling time is shorter than the sound crossing time) and upon compression and rebound breaks up into a central piece that contains most of the mass and several satellites -- a very gentle "shattering" process \citep{mccourt18,gronke20-mist,farber23}. \autoref{fig:static_cloud} shows a snapshot of the simulation box after the configuration is stabilized. The central piece has the largest mass\footnote{Visually, all four pieces are of comparable size. However, the central piece has a large, very dense, cold ($T\sim 10$K) component. If the entire central cloud was at higher temperatures, like the satellite pieces, it would be much larger.} and a large cold ($T \sim 10$K) component. By contrast, the satellite pieces are dominated by $T \sim 5000-10^4$K gas, despite having plenty of time to cool. We zoom in on the central piece and one of the satellite pieces and plot their $v_{\rm turb}$ and Damk$\ddot{\rm o}$hler number. While the central mass is roughly stationary, the satellites are coagulating toward the central mass. This motion, as well as cooling-induced cloud pulsations, drive turbulence, and $v_{\rm turb}$ for the satellite cloud is about an order of magnitude higher than that of the central cloud. As a result, the Damk$\ddot{\rm o}$hler number of central cloud is an order of magnitude above the satellite cloud. This is consistent with their phase structure. The temperature PDF of the satellite cloud is comparable to that of entrained clouds, which have comparable Damk$\ddot{\rm o}$hler number, while the central cloud has much more cold molecular-temperature gas.  Thus, gas motions driven by coagulation alone (without an external wind or turbulent driving) can be sufficient to keep gas at elevated temperatures -- a vivid demonstration of how fragile molecular-temperature gas can be. It also shows that phase structure changes continually with \Da number; Da$\sim 1$ is not an abrupt dividing line. There is a noticeable difference in the temperature PDF between ${\rm Da} \sim 10$ and ${\rm Da}  \sim 100$ gas, and gas must have very high Da to all collapse to the lowest molecular temperature.

\begin{figure*}
\centering
\includegraphics[width=\textwidth]{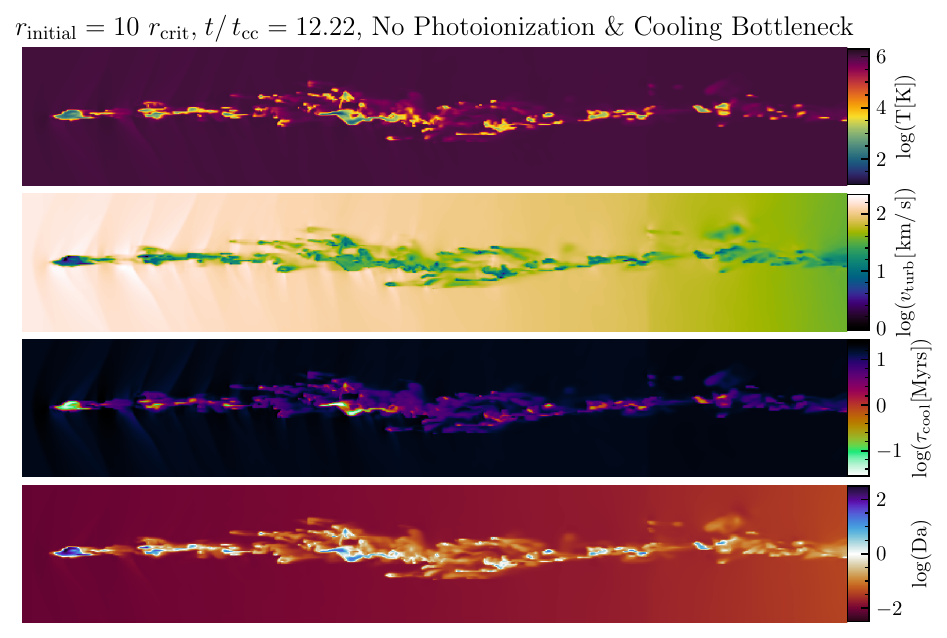}
\caption{Similar to \autoref{fig:Damkohler_number_medium_cloud_during_entrainment}, but for a cloud with $r_{\rm initial} = 10 r_{\rm crit}$, no photoionization, and no cooling bottleneck. Again, the snapshot analyzed here is during cloud entrainment ($\left. t \right/ t_{\rm cc} = 12.22$), which is when the fate of the cloud gets decided. To find L, the characteristic lengthscale, we use a clump finder to identify all cloud fragments and use the average transverse dimension of the clumps as L. The \Da number here is is similar to the \Da number of the fiducial cloud {\it after} entrainment. Indeed, this cloud form a similar amount of molecular-temperature gas and maintains rough mass equipartition between the atomic and molecular-temperature phases. This leads to a cloud overdensity of $\chi \sim 300$.}
\label{fig:Damkohler_number_no_self_shielding_bottleneck}
\end{figure*}

These results also help us understand the relatively low overdensity of the "no self-shielding and cooling bottleneck" cloud {\it during} entrainment, which enables it to entrain despite the fact that it is no longer fully atomic. Although this is an artificial setup which will not be realized in nature, it is helpful in understanding the fundamental robustness of multi-phase cloud entrainment. \autoref{fig:Damkohler_number_no_self_shielding_bottleneck} shows temperature slices and Damk$\ddot{\rm o}$hler number\footnote{Comparing the temperature slice in the top panel of \autoref{fig:Damkohler_number_no_self_shielding_bottleneck} with that of our fiducial simulation shown in \autoref{fig:Damkohler_number_medium_cloud_during_entrainment}, we see a striking difference in cloud morphology. The cloud without photoionization and the cooling bottleneck appears to be much more fragmented. To accurately capture this fragmentation in our choice of L, the characteristic lengthscale, we use a clump finder to identify all cloud fragments and use the average transverse dimension of the clumps as L. Using this method, we found L to be 25.5pc.}  for this case during cloud entrainment, when the fate of the cloud is determined. The \Da numbers of this case is similar to the \Da numbers of our fiducial cloud {\it after} entrainment, when substantial amounts of molecular-temperature gas form. Thus, this cloud also forms similar amounts of molecular-temperature gas as entrained clouds. However as previously noted, in this Da range, there is rough mass equipartition between atomic and molecular-temperature gas, and a broad range of intermediate temperature gas, leading to an overdensity of $\chi \sim 300$ in both cases. 

Note that after entrainment, as turbulence decays, the \Da number, molecular-temperature gas fraction and overdensity $\chi$ all rise. But the rise is slow, even for the "Neither" case, and still only reaches a value of $\chi \sim 1000$ at $t/t_{\rm cc} \sim 50$. Ultimately, the robust low $\chi$ and entrainment in this case comes from the broad temperature PDF which develops even at nominally high values of \Da number in this temperature range, allowing relatively high temperature (low density) gas to dominate the volume, which drives down the mean overdensity. Unlike in the $10^4 \, {\rm K} < T < 10^6 \, {\rm K}$ range, where the temperature PDF is very strongly bimodal, and cooling gas quickly collapses to the $T \sim 10^4$K temperature floor, bimodality is less pronounced in the $10 \, {\rm K} < T < 10^4$K range, likely due to the different shape of the cooling function. We currently lack a fundamental theory for how the temperature PDF changes with the \Da number when Da$>1$, which will be the subject of upcoming work (see \citealt{tan21-lines} for models of the temperature PDF in the ${\rm Da} \ll 1$ regime). However, as noted previously, our numerical results are consistent with high-resolution simulations of turbulent multi-phase gas in the ISM context \citep{Audit:2005,vázquez-semadeni:2012,Saury:2014}, where the lack of discrete phase structure has also been noted.

\subsection{Survival Criterion for Clouds with a Molecular-Temperature Component} \label{sec:Survival Criterion for Clouds with Molecules}

The key insight so far has been that the entrainment of molecular-temperature clouds is essentially identical to the well-studied case of atomic cloud due to large amounts of atomic gas generated via mixing during cloud entrainment. {\it As a practical matter, the survival criterion for multi-phase clouds is the same as well-studied survival criteria of purely atomic clouds.} The cloud is mostly atomic when it entrains, and then cools to molecular temperatures post entrainment. However, it is also interesting to understand how accurately existing survival criteria capture the underlying physics. One test is to see how they fare when extended to a broader temperature range -- e.g., when there {\it is} a significant amount of molecular-temperature gas present during entrainment. 

A brief summary of existing survival criteria is appropriate. The original paper of \citet{GronkeOh:2018} found that clouds can survive if $t_{\rm cool,mix} < t_{\rm cc}$, where $t_{\rm cool,mix}$ is the cooling time of mixed gas at $T_{\rm mix} \sim \sqrt{T_{\rm c} T_{\rm hot}}$, the geometric mean of the hot and cold gas temperatures  \citep{begelman90}. They reasoned that if mixed gas can cool faster than it is produced, then the cloud should survive. This works for the temperature floor of $T \sim 4 \times 10^4$K that \citet{GronkeOh:2018} imposed, but becomes less accurate at lower temperatures. \citet{farber22} rectified this by developing new survival criteria for cloud gas with at $T \sim 1000$K temperature floor. They found that cloud gas can survive if $t_{\rm cool, minmix} < t_{\rm cc}$, where $t_{\rm cool, minmix}= t_{\rm cool}(T_{\rm minmix})$, the isobaric cooling time at $T_{\rm minmix} = \sqrt{T_{\rm wind} T({\rm min}(t_{\rm cool}))}$, where $T_{\rm wind}$ is the wind temperature and $T({\rm min}(t_{\rm cool}))$ is the temperature when $t_{\rm cool}$ reaches its minimum value. This makes physical sense, since gas tends to accumulate at the minimum of the cooling time, which is the usual thermally stable state (as we see from the abundance of atomic gas in our simulations), and this is the gas that is mixing with the wind. Another nuance is the appropriate mixing time. There are another group of papers, which argue that the correct mixing time is not given by the cloud crushing time, but the shearing time of the wind $t_{\rm shear} = r/v_{\rm wind}$ \citep{sparre20,li20,Abruzzo:2023}. \citet{kanjilal21} showed that part of the discrepancy stems for different definitions of "survival": whether cold gas is present after many cloud-crushing times ($t_{\rm cool,mix}$ appropriate) or whether original cloud material survives ($t_{\rm shear}$ appropriate). In the latest iteration in the cloud survival literature, \citet{Abruzzo:2023} drew together a number of these ideas and showed that $t_{\rm cool,minmix}/(\alpha t_{\rm shear}) \lsim 1$ (where $\alpha \sim 7$ is a numerically calibrated paramater) fit an large suite of numerical simulations well\footnote{The numerically calibrated factor $\alpha \sim 7$ renders $t_{\rm mix}$ comparable to $t_{\rm cc} \sim \chi^{1/2} t_{\rm shear}$ for $\chi \sim 100$, but their criteria fares better at higher overdensities.}.

\begin{figure*}
\centering
\includegraphics[width=\textwidth]{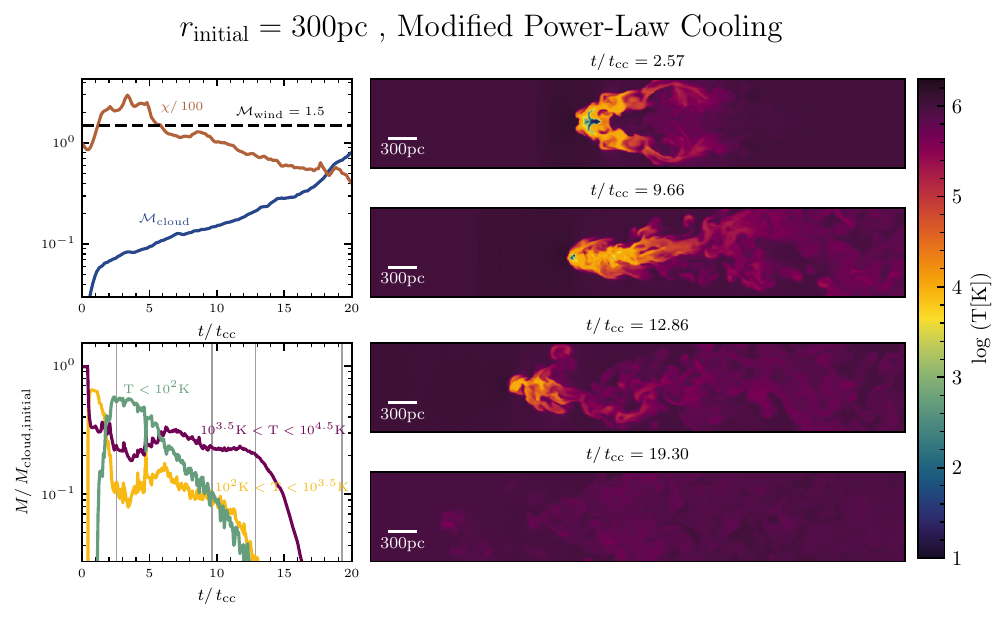}
\caption{Evolution of a cloud with initial size $r_{\rm initial} = 300 {\rm pc}$ and the modified power-law cooling time profile. As we have seen in \autoref{fig:medium_cloud}, the same cloud with all else held equal but with the standard cooling time profile survived, but the four temperature slices plotted in the right column of this figure shows that this cloud with the modified power-law cooling time profile gets destroyed.}
\label{fig:medium_cloud_modified_power_law_cooling}
\end{figure*}

How do such criteria fare when clouds are {\it not} predominantly atomic during the entrainment phase -- so that it does not simply reduce to the canonical case? We have already engineered one such case: when we remove photoionization and cooling bottleneck, molecular-temperature gas is present throughout entrainment. The cloud still entrains and survives. The \citet{farber22} and \cite{Abruzzo:2023} criteria are  equivalent to $r_{\rm crit} \sim v_{\rm wind} t_{\rm cool,minmix}/\sqrt{\chi}$ and $r_{\rm crit} \sim v_{\rm wind} t_{\rm cool,minmix}/\alpha$ respectively. For the cloud with no cooling bottleneck, $T({\rm min}(t_{\rm cool})) \sim 1.8 \times 10^4$K remains the same (see \autoref{fig:cooling_curve}), and thus $t_{\rm cool,minmix}$ is the same. During entrainment, the measured overdensity of the cloud in simulations is $\chi \sim 300$. Thus, $(t_{\rm cool,minmix}/t_{\rm cc}, t_{\rm cool,minmix}/\alpha t_{\rm shear}) \sim (0.04,0.1)$ respectively, which predicts survival, as indeed happens in the simulation. We now wish to construct a case where clouds are destroyed. Consider the "Power-Law Cooling" curve in \autoref{fig:cooling_curve}. This is intended to mimic the cooling time profile between $T\sim 10^4-10^4$K, but now extended down to $T \sim 10$K. In particular, ${\rm min}(t_{\rm cool})$ is the same, but now is changed to be at $T({\rm min}(t_{\rm cool})) \sim 30$K, and there is a sharp rise in cooling time below $T({\rm min}(t_{\rm cool})) \sim 30$K, similar to the sharp rise in cooling time below $T \sim 10^4$K for the standard cooling curve . We run a simulation with $r_{\rm initial} =300 {\rm pc}$ (i.e., the same size as $r_{\rm cl}= 10 r_{\rm crit}$ for the standard cooling curve), and all other wind parameters as before. The evolution is depicted in \autoref{fig:medium_cloud_modified_power_law_cooling}. The cloud is destroyed, despite a relatively low overdensity $\chi \sim 100$ during evolution. Do current survival criterion predict this? In this case, $T_{\rm minmix} \sim \sqrt{30 \times 10^6} \sim 5500 \, {\rm K}$, near the peak of the cooling time: from \autoref{fig:cooling_curve}, $t_{\rm cool}(T\sim 5000 \, {\rm K}) \sim 100 t_{\rm cool}(T\sim 10^5 \, {\rm K})$, where $T_{\rm minmix} \sim 10^5$K for the standard cooling curve. The overdensity of the cloud is $\chi \sim 100$ during entrainment, so $(t_{\rm cool,minmix}/t_{\rm cc}, t_{\rm cool,minmix}/\alpha t_{\rm shear}) \sim (7,10)$ respectively, which predicts destruction, as indeed happens. 

We have a sample size of two, so we would not read too much into these successes (though they are comforting!). The essential intuition behind existing criteria -- that cooling must beat mixing -- is likely correct. But in detail, existing criteria characterize a broad distribution of mixing and cooling times by a single number for each, and thus are simplifications. We have not invested effort into refining cloud survival criteria. Our pragmatic view is that for standard cooling curves, our results show molecular-temperature cloud entrainment is essentially equivalent to atomic cloud entrainment, and there is already a very extensive literature on the latter, with well-validated survival and entrainment criteria. Our point is merely that for those who aspire to a deeper understanding of mixing and entrainment physics (e.g., in the vein of \citealt{Abruzzo:2022}), such numerical experiments -- across a broader temperature range, where one changes cooling curve properties to test hypotheses -- can be very useful in validating theoretical models.  

\section{Discussion}
\label{sec:discussion}

The survival and entrainment of dust and molecules in outflows has long thought to be theoretically puzzling \citep{ferrara16}. In this paper, we have shown that it arises from the large amounts of atomic gas which forms via mixing, shields molecules and dust, and reduces the cloud overdensity. In this section, we put our work in the context of existing literature and discuss its implications. 

\subsection{Observations of Molecular Gas and Dust} \label{sec:observations of molecular gas and dust}

Molecular gas and dust have been observed in and around galaxies at low and high redshifts. We review some pertinent observations here.

Within our own galaxy, molecular cloud observations are often associated with the Milky Way's nuclear wind. The proximity of these clouds allow observers to map out their morphology and make statements about their interaction with the surrounding hot wind. For example, \cite{Di_Teodoro:2020} used CO emission lines to map out the structure and velocity of two molecular clouds at a galactic latitude of $\sim 5^{\circ}$ (which corresponds to a height of 0.7 kpc above the disk). The cloud closer to the disk is more clumpy, while the cloud further away is more filamentary and elongated. They interpret this as direct evidence of cloud-wind interaction, with the elongated cloud in a later evolutionary stage of being disrupted by the hot wind. Given their column density and kinematic measurements, \cite{Di_Teodoro:2020} found the very large amount of high-velocity molecular gas to be puzzling given available acceleration mechanisms, and suggested in-situ molecular gas formation during cloud entrainment and mixing as a solution, as our paper explores. \cite{Cashman:2021} and \cite{Noon:2023} also observed molecular gas in Milky Way's nuclear wind at $< 1$ kpc above the disk. In particular, \cite{Noon:2023} mapped out both the atomic (using 21 cm H line) and molecular (using CO emission line) component of three clouds and found that the clouds are out of chemical equilibrium: the HI column density of these clouds are too low to shield the molecular component in the interior from dissociating. Such non-equilibrium abundances could arise if the clouds are dynamically interacting with the wind. The cloud morphology they observed (Figures 1 and 2 in their paper) has similarities with our simulation results in \autoref{fig:medium_cloud}, \autoref{fig:small_cloud}, and \autoref{fig:large_cloud}, including the presence of an atomic tail. Molecular gas is also seen in high velocity clouds (e.g., \citealt{Tchernyshyov:2022} observed ${\rm H_{\rm 2}}$ absorption in a HVC toward the LMC). 

Looking at the CGM of other galaxies allows us to better understand the prevalence of molecular gas and dust and how they evolve over cosmic times, even though such observations are of course much less well resolved. At low redshifts, dust in the CGM is often detected by its reddening and extinction effects on the spectra of distant quasars and high-redshift galaxies. \cite{Ménard:2010} found dust reddening effects by correlating the brightness of $\sim 85,000$ quasars with the position of 24 million galaxies at z$\sim 0.3$. They detected dust at galacto-centric distances ranging from 20 kpc to several Mpc. \cite{Peek:2015} observed dust reddening in the spectra of distant galaxies due to foreground galaxies at z $\sim 0.05$, out to 150 kpc from the foreground galaxy. Reddening had a power-law decline with impact parameter, indicating more dust closer to the galaxy. Other methods of detecting dust in the CGM of low redshift galaxies include far-infrared emission from cool dust \citep{Roussel:2010}, scattering of UV light by dust grains \citep{Hodges-Kluck:2014}, and using metals as a tracer for dust \citep{Peeples:2014}. In all these cases, dust is observed at least up to 20 kpc away from the galaxy. At high redshifts, molecular gas and dust is mapped by their emission features. \cite{Ginolfi:2017} used ALMA to observe a star-forming galaxy Candels-5001 at ${\rm z} = 3.47$. They traced an elongated molecular gas structure that extends out to 40 ${\rm kpc}$ by CO emission, accompanied by a continuum thermal dust emission at the same spatial scale. 

Molecular gas, dust and star formation is also seen in cluster filaments \citep{sparks09,donahue00}. The spectacular filaments in NGC 1275 in the Perseus cluster are perhaps the best known example. The very large molecular masses (which can reach $\sim 10^{11}\, M_{\odot}$) and high star formation rates, even when the associated BCG is no longer forming stars, suggests that molecular mass growth has decoupled from the BCG. Consistent with the models in this paper, molecular and dusty gas is invariably accompanied by atomic gas. For instance, there is a one-to-one match between BCGs that have H$\alpha$ emission and infrared line emission \citep{donahue00}, and observed H$\alpha$ morphology shows a close correspondence with infrared images of ${\rm H}_2$ emission. More surprisingly, polycyclic aromatic hydrocarbons (PAHs), which are more easily destroyed than dust grains, are also seen, with emission features similar to that in the ISM of star forming galaxies \citep{Donahue:2011}. This suggests that direct mixing with hot ICM gas has been strongly suppressed, consistent with our findings. The existence of dust and molecules in cluster filaments is an important clue in the ongoing debate about whether cold gas forms via precipitation (i.e., thermal instability) or uplift and further condensation onto cold gas originating from the BCG  \citep{donahue22}. While both processes are likely to be at play, at least some cold gas with embedded dust must originate from the BCG. Dust is needed to catalyze molecule formation and seed further grain growth, and dust grains cannot survive for long in the very hot ICM gas; it is unlikely to be present in cool gas that forms via precipitation out of the hot phase. Dust and molecules are also seen in  the tails of "jellyfish" galaxies, which are undergoing ram pressure stripping \citep{moretti18} in galaxy clusters. The jellyfish galaxies have a normal amount of molecular gas, consistent with simulations that find that ram pressure stripping of molecular gas is inefficient \citep{tonnesen10}. At the same time, the molecular tails have masses comparable to the disk, suggesting that the molecular phase formed in situ.

One environment where {\it direct} mixing between molecular and hot gas has been claimed is in the cometary tail of Mira, an AGB star whose molecular wind is seen in far UV emission \citep{martin07}. These authors invoke an unusual emission mechanism whereby $H_{2}$ molecules are collisionally excited by hot electrons from shocked $T \sim 5 \times 10^5$K gas. However, this hypothesis has not yet been reproduced in more detailed calcuations\footnote{Recent hydrodynamical simulations of AGB winds do not include molecular chemistry, but they {\it do} incorporate cooling down to 300 K \citep{li19}. In this study, the cold AGB wind (originally at a few hundred K) quickly heats up to $T \sim 10^4$K -- probably due to mixing-- and stays there.}, and other explanations for the UV emission, such as hydrogen two photon emission or coronal line emission, are also possible.

In summary, dust and molecules are seen in a wide variety of environments, even settings with an abundance of hot gas where mixing processes and thermal sputtering might be expected to destroy them. Although the primary scenario we have in mind while designing our simulation is that of cold clouds being accelerated by hot galactic winds, we have intentionally kept the simulation setup general such that our results are broadly applicable to a wide range of astrophysical systems, some of which we surveyed in this section. The model presented in this paper -- in particular the notion that atomic gas produced by mixing "shields" molecular gas from direct mixing -- potentially explains the survival and growth of dust and molecules in these environments. However, more detailed follow-up work specific to these settings is needed. 

\subsection{Comparison with Previous Work} \label{sec:previous works}

In this section, we put our results in the context of existing theoretical work.

The work most closely related to this paper is \citet{farber22}, a pioneering paper studying the long-term survival of molecular-temperature gas and dust in a wind. The biggest difference is that they set a temperature floor\footnote{They also initialize their clouds at $T \sim 1000 \,$K, whereas we initialize our clouds at $T \sim 10^4$K, though the self-shielded portions of our clouds rapidly cool to molecular temperatures at the beginnning of the simulation, $t \lsim t_{\rm cc}$, which significantly reduces this difference.} of $T \sim 1000 $ K, whereas our gas can cool to $T \sim 10 \, {\rm K}$, allowing greater dynamic range and realism. 
We also include photoionization/self-shielding, and model dust evolution differently. They assume dust (which catalyzes molecule formation) is required to cool to low temperatures, whereas we assume metal fine-structure cooling can operate and bring gas to low temperatures regardless of dust abundance. \citet{farber22} find molecular-temperature gas and dust can survive and entrain in some portions of parameter space, and that abundant atomic gas can form via mixing, both in agreement with our conclusions. However, our respective pictures for when and how this happens differ. \citet{farber22} identified three evolutionary paths for cold clouds, in order of increasing cloud size: (i) destruction, (ii) transformation into $10^4$ K gas, or (iii) survival of the cold phase; they formulated criteria for these 3 different pathways (see \S\ref{sec:The Damkohler Number and Atomic Gas Production Through Mixing}). We do not see significant entrainment of initially molecular-temperature gas, but find that strong mixing during entrainment renders our clouds mostly atomic and low overdensity\footnote{Although, consistent with their results, we do see the molecular-temperature component survives for longer with increasing cloud size.}. Instead, the molecular-temperature component grows after entrainment. 
Due to differences in setup, assumptions, and simulation run-time, \citet{farber22} do not find this "Phoenix"-like behavior, and consequently molecular-temperature gas is significantly more fragile in their simulations.   

Our work is well-complemented by the study of in situ formation of molecular gas by \cite{Girichidis:2021}. Simulations ran by \cite{Girichidis:2021} subject an atomic temperature cloud to a hot, $10^6$K wind. They choose several initial cloud temperatures such that pressure equilibrium is maintained and the initial density contrast is varied from 100 to 1000. Their simulations include many physical processes we neglected, including a magnetized wind, a chemical network that track changes in thermal energy, ${\rm H}_2$ formation on dust grains, self-shielding, a cosmic ray heating term, a turbulent velocity field, and self-gravity. However, they assume a constant dust-to-gas mass ratio (no dust formation or destruction), while we implement both the sputtering and accretion of dust explicitly in our passive scalar model for dust. Another crucial difference is that all of the simulations ran by \cite{Girichidis:2021} focus exclusively on the initial evolutionary stages of the clouds and only evolve them for $\lesssim 10 t_{\rm cc}$ (the densest cloud is only evolved for less than 3$t_{\rm cc}$). Although this approach is useful for studying the in situ formation of molecular gas, the clouds are far from being entrained at the end of their simulations, and it is not clear whether the clouds will eventually survive or get destroyed.

More recently, \cite{Zhang:2023} ran three-dimensional MHD simulations to study whether high-velocity molecular clouds observed in the Milky Way can be accelerated by starburst. They simulate a long box perpendicular to the galactic disk that includes a uniform magnetic field, a physically motivated model for gravitational potential, and a series of random supernova explosions that accelerate a molecular cloud. They find that magnetic fields provide protection and acceleration to their molecular clouds, and the clouds can reach latitudes of $\sim$1kpc with velocities of 200$\left. {\rm km} \right/ {\rm s}$, which are consistent with observations. However, they do not implement a frame-tracking scheme for the cloud, which means they can only consider cloud evolution up to a latitude of 1kpc above the disk. Much like \cite{Girichidis:2021}, \cite{Zhang:2023} do not track the long-term evolution of their molecular clouds and thus are agnostic as to the eventual survival and entrainment, which is needed to explain observations of molecular gas and dust far out in the CGM.

\subsection{Caveats and Missing Physics} \label{sec:caveats}

In this section, we identify some physical effects that are not fully captured in our simulations, which are possible directions for future work. 

Broadly speaking, our work shows that the entrainment of molecular-temperature gas is very similar to the entrainment of purely atomic gas, since abundant atomic gas is formed by mixing, and the resulting cloud has overdensities comparable to purely atomic clouds. Furthermore, hot wind gas seldom directly interacts with cold molecular-temperature gas, but instead with warm $T \sim 10^4$K gas. Thus, our strengths and failings are very similar to that of classic hydrodynamic wind tunnel simulations of atomic gas clouds, which were recently reviewed in detail in \citet{Faucher-Giguere:2023}. There is more discussion there of the effects of physics we have omitted such as thermal conduction, magnetic fields, cosmic rays, turbulence, and more realistic (spatially and temporally evolving) winds on cloud entrainment and growth. Very briefly: although B-fields strongly suppress mixing in planar shear mixing layers \citep{ji19,zhao23}, they do not seem to significantly change mass growth and entrainment rates for clouds \citep{gronke20-cloud,li20,sparre20,jennings23}, for reasons that are not yet fully understood\footnote{They do, however, have a drastic impact on cloud morphology, making it much more filamentary.}. Thermal conduction similarly has mild effects, in part because once magnetic fields drape the cloud, anisotropic conduction is suppressed \citep{li20,jennings23}, and turbulent heat diffusion tends to be stronger than thermal heat conduction \citep{tan21}. Cosmic rays can contribute to cloud acceleration via the "bottleneck effect" \citep{wiener17-cold-clouds,wiener19,bruggen20,bustard21,huang22-clouds}, though there has not been a detailed study of how they affect cloud entrainment in a pre-existing wind. Turbulence can cause the cloud to break up and form a power-law mass distribution \citep{gronke22-turb,TanFielding:2023}, though the entire complex still grows in mass and has strong momentum coupling to the hot gas. Higher temperature (which implies higher overdensity $\chi$; see Appendix \ref{sec:Twind1e7K simulation}) and Mach number winds can also have effects which are challenging to simulate, and deserve more exploration. In particular, 3D simulations are not converged at high Mach numbers \citep{gronke20-cloud}; only 2.5D simulations are well converged \citep{bustard22}. A final twist is the case of infall, rather than outflow, where gravity conspires to create unrelenting shear, resulting in a drag-induced terminal velocity, and a different survival criterion $t_{\rm grow} < t_{\rm cc}$ from the wind tunnel case \citep{tan23-gravity}. 

Caveats more specific to this work are related to our treatment of photoionization, dust, and cooling to lower temperatures $T < 10^4$K. Our treatment of photoionization and self-shielding is fairly simplistic, but our conclusions are likely not sensitive to this, given that clouds robustly entrain and grow even if we do not model a photoionized "skin" (\autoref{sec:What Role Does Photo-ionization and the Cooling Bottleneck Play?}). We also do not model the dependence of radiative cooling on the radiation field, metallicity \footnote{We assume solar metallicity throughout, whereas it is likely that the CGM is lower metallicity, so our clouds will gradually decrease in metallicity as they mix and grow in accreted mass.} and/or dust or molecular-temperature content. We justify the last approximation by the dominant contribution of metal fine-structure line cooling at low temperatures \citep{Krumholz:2012,Glover:2013}. Our single-size ($a \sim 0.1 \, {\mu}$m) treatment of dust is also fairly approximate. While the assumption of perfect coupling with gas is in line with other treatments using Lagrangian tracer particles \citep{silvia10,farber22}, it ignores possible inertial sputtering due to the relative motion of dust and gas \citep{hu19}. We also do not explicitly track the formation or possible photo-dissciation of molecules. However, our clouds experience a order of magnitude growth in mass and form dusty, low temperature, high density interiors. Thus, the formation of molecules is expected to proceed efficiently with dust grains as catalyst, and self-shielding to photo-dissociating radiation is likely. In future work, this should be treated explicitly.  

We also note that we ignore self-gravity of the cloud, which is fine in atomic pressure-confined clouds, but may become important in cold molecular clouds. To get a sense of whether self-gravity is important in our simulations, it is instructive to compare the Jeans mass with the mass of our cloud in different temperature phases. The Jeans length is given by

\begin{align}
    \lambda_{\rm J} \sim c_{\rm s,eff} t_{\rm dyn} = (c_{\rm s}^2 + v_{\rm turb}^2)^{1/2} \frac{1}{\sqrt{{\rm G} \rho}}, \label{eq:Jeans_length}
\end{align}

where we have defined an effective sound speed $c_{\rm s,eff} = (c_{\rm s}^2 + v_{\rm turb}^2)^{1/2}$ to account for the role of turbulent pressure support in opposing collapse under self-gravity. As shown in \autoref{sec:understading molecular cloud entrainment}, $v_{\rm turb} \sim c_{\rm s}$ both in the atomic phase during entrainment and in the cold molecular-temperature phase after entrainment. 


We compare the temperature and density dependent Jeans mass $M_{\rm J}$ with the cloud mass in our fiducial simulation with $r_{\rm initial} = 10 r_{\rm crit} = 300 {\rm pc}$ for two different phases: the atomic ($T \sim 10^4$K) phase and the cold molecular-temperature phase ($T \lesssim 100$K). 
For the atomic phase, $\left. M_{\rm J,10^4K} \right/ M_{\rm initial} \sim 3 \times 10^4$. Since the mass of the atomic phase never exceeds $10 M_{\rm initial}$ (see the bottom left panel of \autoref{fig:medium_cloud}),  
self-gravity is unimportant in the atomic phase. As the cloud is primarily atomic during entrainment, our main conclusions about cloud survival and efficient entrainment should be unaffected by self-gravity. 
For the cold molecular-temperature phase, $\left. M_{\rm J,<100K} \right/ M_{\rm initial} \sim 3$; note that the cold molecular-temperature phase grows to $\sim 10 M_{\rm initial}$ after entrainment. 
Although (due to fragmentation) individual cold clumps in our simulations remain below the Jeans mass, molecular clouds could certainly become self-gravitating in some portions of parameter space.


Indeed, star formation is seen in outflows \citep{maiolino17}, as well as the tails of "jellyfish" galaxies \citep{moretti18} 
However, the crucial point for the purposes of this manuscript is that clouds are mostly atomic before entrainment, and hence self-gravity does not affect the entrainment process, the main focus of this paper. Once the cloud is entrained and molecular gas forms, then indeed self-gravity can become important and result in fragmentation and star formation. 
The fate of self-gravitating cold molecular-temperature gas in outflows is certainly an interesting open question, but it is beyond the scope of this paper.


We also note that ISM scale simulations \citep{Girichidis:2016} found that the outflows generated from the ISM are initially atomic, although they include both self-gravity and molecular chemistry. This is consistent with our findings, since they have a relatively short run time, consistent with the purely atomic entrainment phase in our simulations, before molecular gas appears. Also consistent with the above, they find that self-gravity is important in the ISM, but not in the wind.

Additionally, we do not include extrinsic turbulence in our simulations. Turbulence promotes cloud fragmentation and results in a power-law distribution of cloud sizes, as seen in both idealized \citep{gronke22-turb} and galaxy-scale \citep{tan23-cloud-atlas} simulations. The clouds eventually trace the velocity field of the hot gas, with similar structure functions \citep{gronke22-turb}. However, cloud survival criteria and total mass growth rates are relatively unchanged \citep{gronke22-turb}, and atomic cloud entrainment is similarly robust \citep{tan23-cloud-atlas}. We therefore tentatively expect that extrinsic turbulence will not significantly change our results, though of course verification would require further work beyond the scope of this paper. 

While additional refinements can and should be pursued -- indeed, the complementary work of \citet{Girichidis:2021} incorporates much of the physics such as molecular chemistry, magnetic fields, self-gravity, etc that we elide -- we do not believe these omissions affect our major conclusions.  

\section{Summary} \label{sec:summary}

In this work, we present 3D hydrodynamic simulations of molecular-temperature clouds in a hot wind. We include prescriptions for photoionization and self-shielding. Thus, our clouds develop an optically thin atomic ($\sim 10^4$ K) skin that shields the cloud interior from both photoionization and mixing. We allow the optically thick interior of the clouds to cool down to molecular temperatures. Dust evolution is tracked using a passive scalar that accounts for both sputtering by thermal collisions and mass growth via accretion. We analyze simulation results of a suite of different cloud sizes, $r=3,10,30 \, r_{\rm crit}$, where $r_{\rm crit}$ is the critical radius for cloud survival and entrainment in the purely atomic case. Finally, we run various numerical experiments with modified cooling curves to further understand the physical origin of our results. 

Our key findings are as follows: 
\begin{itemize}
\item{Even though we do not impose any temperature floors (except in the photoionized skin), and cooling times to molecular temperatures are very short ($\lsim 10\%$ of the cloud entrainment time), clouds are mostly atomic\footnote{While we start with a purely atomic clouds, self-shielded clouds (our $r=10,30 \, r_{\rm crit}$ cases) rapidly develop  molecular-temperature cores while still subject to the full force of the wind, so the initial conditions actually correspond to clouds with roughly equal masses in the atomic and molecular-temperature phases.} ($T\sim 10^4$K) during entrainment. While photoionization plays a role, the most important reason for this is strong turbulent mixing. Before entrainment, when shear is large, the mixing time is short or comparable to the cooling time at $T \sim 5000$K, a "cooling bottleneck" which arises due to hydrogen recombination ($t_{\rm mix}/t_{\rm cool} \lsim 1$). Gas cannot cool past this barrier before mixing to higher temperatures.} 
\item{Since gas is mostly $T \sim 10^{4}$K during entrainment, the cloud maintains a relatively low overdensity $\chi \sim 100$, which greatly facilitates entrainment, compared to the $\chi \sim 10^4-10^5$ overdensities one naively expects if gas can cool to very cool molecular temperatures. Survival and entrainment criteria are identical to the well-studied purely atomic case, and all our simulated $r=3,10,30 \, r_{\rm crit}$ clouds fully entrain.}
\item{Any cloud that survives entrainment eventually produces cold molecular-temperature gas ($T \leq 100$K) in its self-shielded interior \emph{after} entrainment, when shear falls and the mixing time increases, so that $t_{\rm mix}/t_{\rm cool} \gsim 1$. The detailed mass evolution of the different phases depends on the initial cloud size. Roughly speaking, clouds that are initially larger generate molecular-temperature gas faster. However, the gas does not all collapse to the minimum molecular temperature $T \sim 10$K. Instead, by the time the entrained cloud has grown by an order of magnitude in mass, there is roughly equal mass in atomic and molecular-temperature phases. Consistent with simulations of turbulent, thermally unstable gas in the ISM literature, gas occupies a broad range of temperatures, including gas in formally thermally unstable range $50\, {\rm K} < T < 5000 \, {\rm K}$. This reduces the overall overdensity of the cloud. The temperature PDF tracks the cooling time as a function of temperature $t_{\rm cool}(T)$, particularly features such as the "cooling bottleneck". We will pursue an analytic model for the temperature PDF in the $t_{\rm mix}/t_{\rm cool} > 1$ regime in future work.}  
\item{We artificially modify the cooling function to change $t_{\rm cool}(T)$. Under these modifications, current cloud survival criteria correctly predict whether clouds are destroyed ("Modified Power Law Cooling"; \autoref{fig:medium_cloud_modified_power_law_cooling}) or survive and entrain  ("Neither Self-Shielding nor Photoionization"; \autoref{fig:self_shielding_and_cooling_bottleneck}; despite a significant molecular-temperature component during entrainment $\chi \sim 300-2000$). These results should not be over-interpreted, given the small sample size, and we emphasize that such modifications do not arise in real world applications. We already have a robust criterion for molecular-temperature cloud entrainment. Nonetheless, such numerical experiments can be very helpful in testing theoretical models which aspire to a deeper understanding of the interaction between mixing and cooling.}
\item{In our simulations, cloud dust mass is roughly constant (within a factor of $\sim 2$ of the initial dust mass), indicating that destruction by thermal sputtering is not dominant. Most dust is protected by $T \sim 10^4$K atomic gas; grain growth can also counteract the loss of the dust by mixing.  However, there is significant dilution of the initial dust mass fraction due to the order of magnitude increase in cloud mass, from cooled dust-free wind gas. Our entrained clouds typically have $\left. M_{\rm dust} \right/ M_{\rm dust,initial} \sim 0.7$ and $\left. \mathcal{D} \right/ \mathcal{D}_{\rm initial} \sim 0.07$. These conclusions are specific to our parameter choices; dust will be able to grow more efficiently in denser environments. Nonetheless, sufficient dust should be present to catalyze molecule formation, which we do not directly simulate.} 
\item{We have checked that our results for mass evolution in different temperature ranges are numerically converged, and that similar results for molecular-temperature gas and dust evolution hold in a hotter  wind ($T \sim 10^7$K; $\chi \sim 1000$ for a clouds at $T \sim 10^4$K).}

Our results are relevant to the highly diverse set of environments -- galactic and AGN winds, "jellyfish" galaxies, cluster filaments -- where dust and molecules are seen, even settings with an abundance of hot gas where
mixing processes and thermal sputtering might be expected to destroy
them. The model presented in this paper – in particular the notion that
atomic gas produced by mixing ‘shields’ molecular gas from direct
mixing– potentially explains the survival and growth of dust and
molecules in these environments. However, more detailed follow-up
work specific to these settings is needed.

\end{itemize}

\section*{Acknowledgements}

We thank the anonymous referee for providing detailed feedback that improved the quality of this manuscript. We thank Brent Tan, Ish Kaul, Navin Tsung, Enrique Vazquez-Semadeni, and Zixuan Peng for useful discussions. We acknowledge support from NASA grant 19-ATP19-0205 and NSF grant AST-1911198. Additionally, this work made considerable use of the Stampede2 supercomputer through allocations TG-PHY210004, TG-PHY230074, and TG-PHY230107 from the Advanced Cyberinfrastructure Coordination Ecosystem: Services \& Support (ACCESS) program, which is supported by National Science Foundation grants \#2138259, \#2138286, \#2138307, \#2137603, and \#2138296.

\section*{Data Availability}

The data underlying this article will be shared upon reasonable request to the corresponding author.



\bibliographystyle{mnras}
\bibliography{reference, master_references}




\appendix

\section{Convergence Test} \label{sec:convergence test}

\begin{figure*}
\centering
\includegraphics[width=\textwidth]{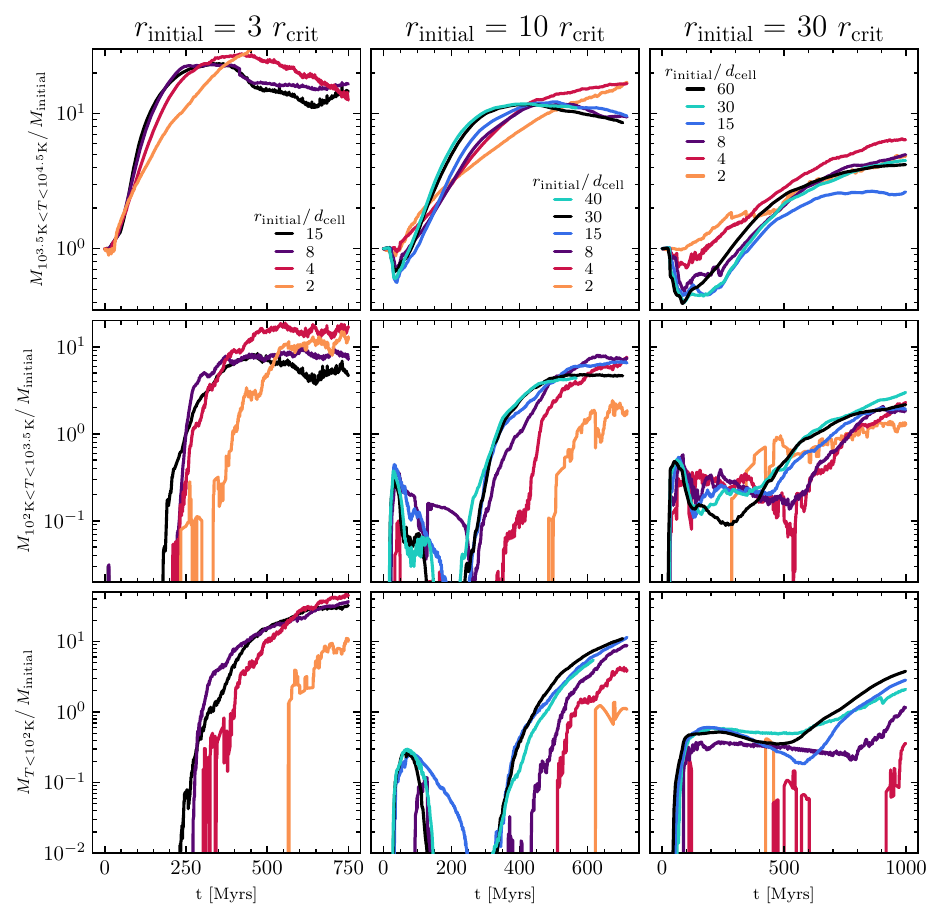}
\caption{Convergence test of our simulations with photoionization. We plot the mass evolution of the small, medium, and large clouds we discussed in \autoref{sec:results} in the left, middle and right column, and the mass evolution in the atomic ($10^{3.5} {\rm K} < {\rm T} < 10^{4.5} {\rm K}$), warm molecular ($10^{2} {\rm K} < {\rm T} < 10^{3.5} {\rm K}$), and cold molecular (${\rm T} < 10^{2} {\rm K}$) temperatures in the top, middle, and bottom row. The black curves represent the fiducial resolution choices we use in the rest of this work. As shown in this figure, the mass evolution of the small, medium, and large cloud converges at around $\left. r_{\rm initial}\right/d_{\rm cell} = $ 8, 15, and 15, respectively, compared to our fiducial resolution of $\left. r_{\rm initial}\right/d_{\rm cell} = $ 15, 30, and 60.}
\label{fig:convergence_test}
\end{figure*}

We tested the convergence of our three fiducial simulations (\autoref{sec:results}) with the standard cooling curve, self-shielding, and $\left. r_{\rm initial} \right/ r_{\rm crit} =$3, 10, and 30. As for the other simulations discussed in \autoref{sec:What Role Does Photo-ionization and the Cooling Bottleneck Play?}, which use a modified cooling curve or do not implement photoionization, we note that: i) the minimum cooling time $t_{\rm cool,min}$ of these simulations is either comparable or larger than our fiducial simulations, albeit shifted to a different temperature, and that convergence in mass growth mainly requires the driving scale of turbulence to be resolved \citep{tan21}. Thus, we do not expect our changes in our cooling curve to affect convergence. (ii) Simulations without photoionization converge at a lower resolution, since they do not require the optically thin atomic skin to be resolved. Thus, we only test the convergence of our three fiducial simulations. The results of our convergence test is shown in \autoref{fig:convergence_test}. The mass evolution of the small, medium, and large cloud converges at around $\left. r_{\rm initial}\right/d_{\rm cell} = $ 8, 15, and 15, respectively, compared to our fiducial resolution of $\left. r_{\rm initial}\right/d_{\rm cell} = $ 15, 30, and 60, respectively. Note that these resolutions correspond to the cloud size at $T \sim 10^4$K. Even though the portions of the cloud interior with $T < 100$K are more poorly resolved due to their small size at high density, their mass evolution is still fairly well converged.

\section{A Simulation with $T_{\rm wind}=10^7$K} \label{sec:Twind1e7K simulation}

The wind temperature in our simulations is set at $T=10^6$K. However, a higher wind temperature is conceivable in astrophysical environments like the intracluster medium (ICM) or in strong starbursts. In this section, we rerun our simulation of the $r_{\rm intial} = 3 r_{\rm crit}$ cloud with a wind temperature of $10^7$K to show that our results are robust and qualitatively unaffected even with higher wind temperatures.

\begin{figure*}
\centering
\includegraphics[width=\textwidth]{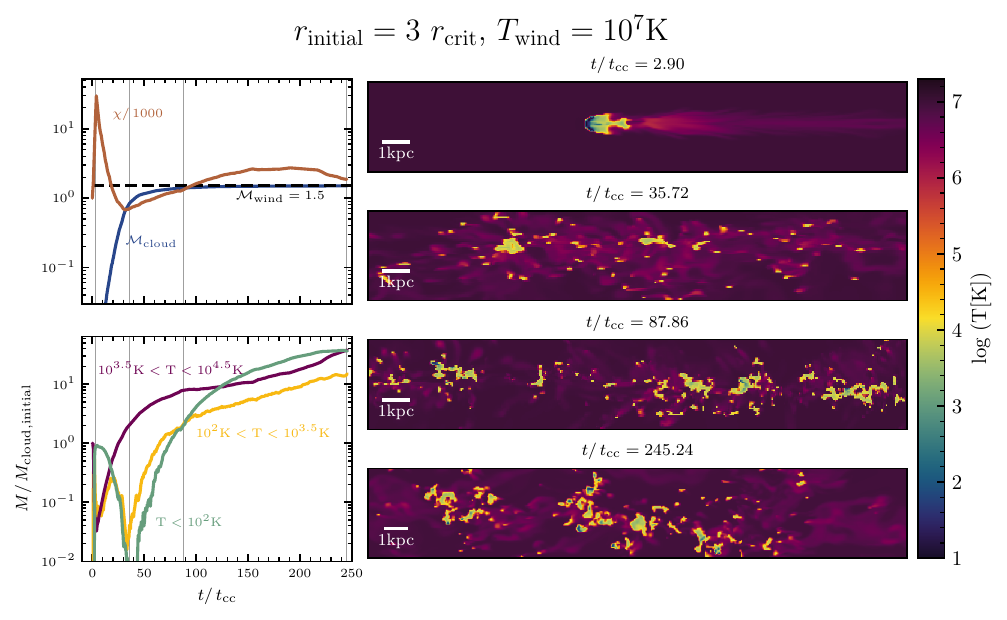}
\caption{Similar to \autoref{fig:medium_cloud}, but with a wind temperature of $10^7$K and no photoionization.}
\label{fig:Twind1e7K}
\end{figure*}

\begin{figure*}
\centering
\includegraphics[width=\textwidth]{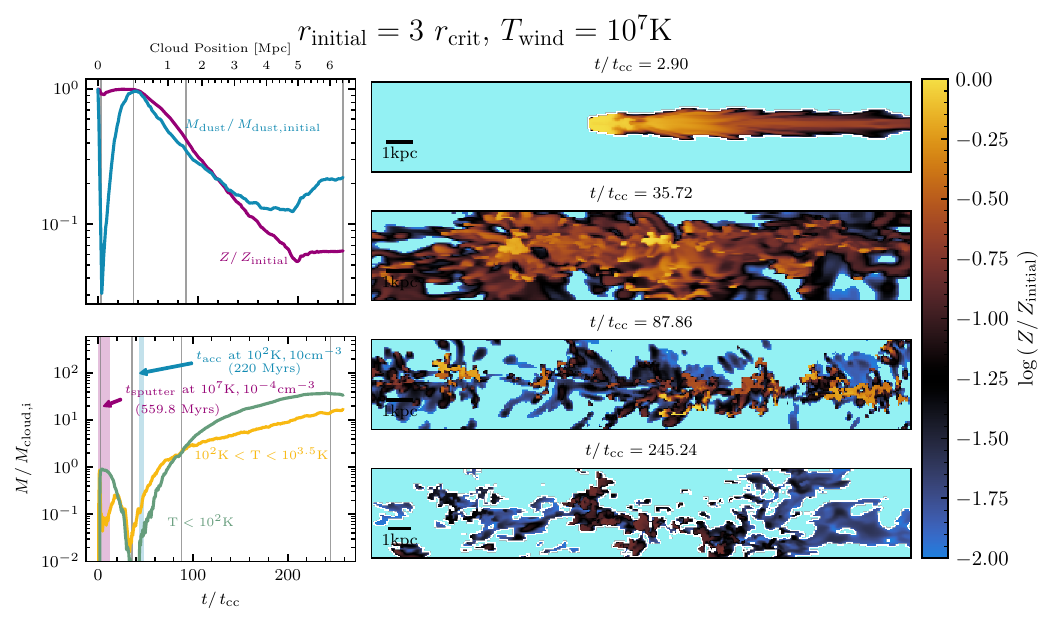}
\caption{Similar to \autoref{fig:medium_cloud_dust}, but with a wind temperature of $10^7$K and no photoionization. }
\label{fig:Twind1e7K_dust}
\end{figure*}

A direct consequence of increasing the wind temperature is an increase in the density contrast $\chi$, which means the critical radius for cloud survival ($r_{\rm crit}$, \autoref{eq:r_crit abruzzo}), cloud entrainment time, and simulation box size (whose length scales as $\sim \chi r_{\rm initial}$) requirements all increase. In view of the additional computational cost, we only simulate a single case, and do not implement photoionization, to avoid having to resolve the optically thin skin before self-shielding kicks in, which increases resolution requirements. We have shown in \autoref{sec:What Role Does Photo-ionization and the Cooling Bottleneck Play?} (see \autoref{fig:self_shielding_and_cooling_bottleneck}) that the inclusion of photoionization does not qualitatively affect the evolution of out clouds.

In \autoref{fig:Twind1e7K} and \autoref{fig:Twind1e7K_dust}, we show the morphology and dust evolution of this cloud. With a longer cloud-crushing time due to the larger density contrast, this cloud has more time to cool and contract in the initial stages of its evolution, as shown by the first snapshot in \autoref{fig:Twind1e7K}. This increases the average density contrast and dust concentration of the cloud significantly before the cloud gets broken apart by the wind. Note also the very significantly different and much more fragmentary morphology of the cloud complex, compared to our fiducial runs. In the $\chi \sim 100$ fiducial case, the cloud "shatters" initially but then re-coagulates when it stabilizes at $\langle T \rangle \sim 10^4$K, $\langle \chi \rangle \sim 100$. In contrast, the significantly higher overdensity of the cloudlets here makes coagulation much more difficult, and the complex remains in a "shattered" state, similar to the "no photoionization or cooling bottleneck" case, which also has high initial overdensities. Molecular-temperature gas is destroyed completely when the cloud breaks apart but quickly reforms as the cloud fragments grow and recombine. Despite the rapid initial contraction and increase in overdensity, the average density contrast of this cloud eventually settles at $\chi \sim 1000$ during entrainment, which is consistent with cloud gas at $T \sim 10^{4}$K in thermal pressure balance with its surroundings. Most of the dust content in original cloud is exposed to the hot wind and gets destroyed during cloud entrainment, but this is compensated by grain growth in the newly formed molecular-temperature phase of the entrained cloud. Although the average dust mass fraction in the cloud dropped to $\sim$ 10\% the original value, dust mass stabilizes and eventually grows in the entrained cloud.

Overall, the mass and dust evolution of this cloud, which includes destruction and revival of molecular-temperature gas, is qualitatively similar to what we discussed in \autoref{sec:intermediate cloud} with a wind temperature of $10^6$K.


\bsp	
\label{lastpage}
\end{document}